\documentclass[journal]{IEEEtran}
\usepackage{amsmath,amsthm,amssymb}
\usepackage{mathrsfs,mathtools,mathabx}
\usepackage{bm}
\usepackage{graphicx}
\usepackage{textcomp}
\usepackage{xcolor}
\usepackage{dsfont}
\usepackage{algorithm,algpseudocode}
\usepackage{algpseudocode}
\usepackage{pgfplots}
\pgfplotsset{compat=newest}
\usepackage{subfigure}
\usepackage{cite}
\usepackage{float}
\usepackage{xparse}
\usepackage[bookmarks=false, hidelinks]{hyperref}
\usepackage{booktabs} 
\usepackage{tikz}
\usetikzlibrary{spy}
\usetikzlibrary{arrows,decorations.pathreplacing,patterns}

\usepackage{etoolbox}

\algnewcommand{\Initialize}[1]{%
  \State \textbf{Initialize:}
  \Statex \hspace*{\algorithmicindent}\parbox[t]{.8\linewidth}{\raggedright #1}
}

\graphicspath{{./Figures/}} 

\makeatother

\newtheorem{prop}{Proposition}
\newtheorem{theorem}{Theorem}	
\newtheorem{rem}{Remark}	

\newtheorem{exam}{Example}
\newtheorem{cor}{Corollary}
\theoremstyle{definition}
\newtheorem{definition}{Definition}

\definecolor{chestnut}{rgb}{0.8, 0.36, 0.36}
\definecolor{airforceblue}{rgb}{0.36, 0.54, 0.66}
\definecolor{cadmiumorange}{rgb}{0.93, 0.53, 0.18}
\definecolor{bleudefrance}{rgb}{0.19, 0.55, 0.91}
\definecolor{carolinablue}{rgb}{0.6, 0.73, 0.89}
\definecolor{blue(ncs)}{rgb}{0.0, 0.53, 0.74}

\definecolor{pearl}{rgb}{0.94, 0.92, 0.84}

\definecolor{asparagus}{rgb}{0.53, 0.66, 0.42}
\definecolor{cssgreen}{rgb}{0.0, 0.5, 0.0}
\definecolor{cadmiumgreen}{rgb}{0.0, 0.42, 0.24}
\definecolor{cadmiumorange}{rgb}{0.93, 0.53, 0.18}
\definecolor{amaranth}{rgb}{0.9, 0.17, 0.31}
\definecolor{bluegray}{rgb}{0.4, 0.6, 0.8}
\definecolor{cadmiumgreen}{rgb}{0.0, 0.42, 0.24}
\definecolor{amaranth}{rgb}{0.9, 0.17, 0.31}
\definecolor{amethyst}{rgb}{0.6, 0.4, 0.8}
\definecolor{amber}{rgb}{1.0, 0.75, 0.0}
\definecolor{azure}{rgb}{0.0, 0.5, 1.0}
\definecolor{babyblue}{rgb}{0.54, 0.81, 0.94}
\definecolor{bazaar}{rgb}{0.6, 0.47, 0.48}
\definecolor{celestialblue}{rgb}{0.29, 0.59, 0.82}
\definecolor{darklavender}{rgb}{0.45, 0.31, 0.59}
\definecolor{bluebell}{rgb}{0.64, 0.64, 0.82}
\definecolor{chamoisee}{rgb}{0.63, 0.47, 0.35}
\definecolor{darkcerulean}{rgb}{0.03, 0.27, 0.49}
\definecolor{iris}{rgb}{0.35, 0.31, 0.81}
\definecolor{dodgerblue}{rgb}{0.12, 0.56, 1.0}
\definecolor{celestialblue}{rgb}{0.29, 0.59, 0.82}
\definecolor{jazzberryjam}{rgb}{0.50, 0.62, 0.37}

\definecolor{burntsienna}{rgb}{0.81, 0.40, 0.26}
\definecolor{burntumber}{rgb}{0.54, 0.2, 0.14}
\definecolor{bulgarianrose}{rgb}{0.28, 0.02, 0.03}
\definecolor{burgundy}{rgb}{0.5, 0.0, 0.13}
\definecolor{cordovan}{rgb}{0.54, 0.25, 0.27}
\definecolor{eggplant}{rgb}{0.38, 0.25, 0.32}
\begin{document}

\title{SumComp: Coding for Digital Over-the-Air Computation via the Ring of Integers

 	\thanks{S. Razavikia and C. Fischione are with the School of Electrical Engineering and Computer Science KTH Royal Institute of Technology, Stockholm, Sweden (e-mail: sraz@kth.se, carlofi@kth.se). C. Fischione is also with Digital Futures of KTH. }
 	\thanks{José Mairton B. da Silva Jr. is with the Department of Information Technology, Uppsala University, Sweden (email: mairton.barros@it.uu.se).}
        \thanks{S. Razavikia was supported by the Wallenberg AI, Autonomous Systems and Software Program (WASP).}
 	\thanks{Jose Mairton B. da Silva Jr. was jointly supported by the European Union’s Horizon Europe research and innovation program under the Marie Skłodowska-Curie project FLASH, with grant agreement No 101067652; the Ericsson Research Foundation, and the Hans Werthén Foundation.}
  \thanks{The EU FLASH project, SSF SAICOM project, the Digital Futures project DEMOCRITUS, and the Swedish Research Council Project MALEN partially supported this work.}
}   

%
\author{
Saeed Razavikia,~\IEEEmembership{Member,~IEEE,}
José Mairton Barros Da Silva Junior,~\IEEEmembership{Member,~IEEE,}
Carlo Fischione,~\IEEEmembership{Fellow,~IEEE}
}

\maketitle

\begin{abstract}
Communication and computation are traditionally treated as separate entities, allowing for individual optimizations. However, many applications focus on local information's functionality rather than the information itself. For such cases, harnessing interference for computation in a multiple access channel through digital over-the-air computation can notably increase the computation, as established by the ChannelComp method. However, the coding scheme originally proposed in ChannelComp may suffer from high computational complexity because it is general and is not optimized for specific modulation categories. Therefore, this study considers a specific category of digital modulations for over-the-air computations, quadrature amplitude modulation (QAM) and pulse-amplitude modulation (PAM), for which we introduce a novel coding scheme called SumComp.   
Furthermore, we derive a mean squared error (MSE) analysis for SumComp coding in the computation of the arithmetic mean function and establish an upper bound on the mean absolute error (MAE) for a set of nomographic functions. Simulation results are presented to affirm the superior performance of SumComp coding compared to traditional analog over-the-air computation and the original coding in ChannelComp approaches in terms of both MSE and MAE over a noisy multiple access channel. Specifically, SumComp coding shows at least $10$ dB improvements for computing arithmetic and geometric mean on the normalized MSE for low noise scenarios. 
\end{abstract}


\section{Introduction}

The evolution of wireless communications, crucial to achieving the omnipresence of the Internet of Things (IoT), has transitioned from human-centric to machine-type communications. The machine learning-based IoT applications in 6G~\cite{tataria20216g}, and the continuous growth of IoT devices~\cite{hellstrom2022wireless,daei2023blind} facilitate handling vast data volumes. To alleviate such heavy communication burden, over-the-air computation (AirComp) has emerged as a solution for simultaneous data collection and computation at the network edge, exploiting the multiple access channel (MAC) waveform superposition property~\cite{goldenbaum2013harnessing,nazer2007computation}.

AirComp is a non-orthogonal wireless communication method that optimizes communication rates and network coverage while preserving energy and data security. Notably, AirComp throughput scales linearly with the number of devices sharing the MAC~\cite{abari2016over}, yielding significant gains for dense networks. Currently a hotbed of research in the wireless community, AirComp has potential applications in distributed machine learning, intra-chip communications, and wireless control~\cite{csahin2023survey}. However, AirComp is limited to analog amplitude modulation. Thus, we recently proposed a new communication for computation method termed ChannelComp~\cite{razavikia2023computing}, which allows the execution of arbitrary finite functions over the MAC and using digital modulations. ChannelComp enhancement broadens the scope of potential applications beyond the mere summation operations supported by the AirComp method.

However, due to its generality, ChannelComp's coding scheme may suffer from computational complexity. Therefore, in this paper, we present a major advancement of the coding scheme for the ChannelComp method while preserving the advantages of low latency and spectral efficiency inherent to analog AirComp and  ChannelComp. This enhancement in the coding scheme overcomes the limitations of analog AirComp's feasibility and ChannelComp coding's complexity and ensures full integration with existing digital communication systems.

\subsection{Literature Review}

The literature reveals two primary AirComp methodologies: uncoded analog aggregation~\cite{gastpar2008uncoded,nazer2007computation} and coded digital AirComp ~\cite{soundararajan2012communicating,wagner2008rate}. The uncoded analog AirComp is extensively studied from various perspectives~\cite{nazer2007computation,goldenbaum2013harnessing,chen2018over,ang2019robust,goldenbaum2013robust} and demonstrates significant resource efficiency, particularly in distributed learning, thus garnering attention for federated edge learning systems~\cite{yang2020federated,amiri2020federated}. The coded digital AirComp approach employs the nested lattice codes' linearity for computation over Gaussian channels~\cite{nazer2011compute,goldenbaum2014nomographic}.

However, AirComp's dependency on analog communication presents reliability challenges due to channel ramifications, e.g., noise and fading effects~\cite{csahin2022over} and the requirements for analog hardware. With its superior channel correction properties and widespread adoption, digital modulation appears more favorable but poses difficulties due to the incoherence of overlapped digitally modulated signals~\cite{zhu2019broadband,razavikia2023computing}.

Recent developments in digital aggregation include methods such as one-bit broadband digital aggregation (OBDA)~\cite{zhu2020one} and majority vote frequency-shift keying (FSK)~\cite{csahin2023distributed}. Moreover, some variants have been proposed, such as the phase asynchronous orthogonal frequency-division multiplexing (OFDM)-based OBDA variant~\cite{you2023broadband}, and non-coherent communication solutions for AirComp~\cite{csahin2022over,csahin2023distributed,sahin2024over}. For instance, in \cite{sahin2024over}, the majority vote function was computed by nullifying a transmitter’s contribution to the superposed value by encoding the votes into the zeros of a Huffman polynomial. These studies work based on the type-based multiple access principle, in which the frequency histogram is estimated for computing the average function by using orthogonal resources for different function output values~\cite{mergen2006type,per2024waveforms}. 

However, these studies mostly focused on specific functions (e.g., sign or summation function) or unique machine learning training processes (e.g., signSGD~\cite{bernstein2018signsgd}). Proposals have used balanced number systems for summation functions, but with increased bandwidth usage~\cite{csahin2022over}. Along this direction, an alternate encoding-based numeral system was introduced~\cite{tang2022radix}. 

Furthermore, the authors in \cite{qiao2024massive} utilize vector quantization to decrease the uplink communication overhead. They apply shared quantization and modulation codebooks specifically to address the federated edge learning problem. This method assumes that the number of transmitting devices is significantly less than the codebook size.

Despite these advancements, the usage of digital modulations with AirComp was limited to nomographic functions~\cite{goldenbaum2014nomographic,goldenbaum2013reliable} and simple digital modulations, such as BPSK or FSK. Indeed, these methods have enforced analog AirComp on digital communication, which was effective only in limited cases and often resulted in inefficient resource usage~\cite{yao2024wireless}.

Recently, a new framework, denoted as ChannelComp~\cite{Saeed2023ChannelComp}, proposed general digital modulations for function computation. This framework aimed at addressing the issue of destructive overlaps of constellation points over the air.  The groundbreaking idea of ChannelComp was to design digital modulation to circumvent destructive overlaps.  In \cite{liu2024digital,xie2023joint}, channel coding techniques are applied based on the same principle to increase communication reliability.

Specifically, ChannelComp represents the constellation points of modulated signal $x$ with a vector $\bm{x} \in \mathbb{C}^{q}$, where $q$ represents the order of modulation. ChannelComp includes constraints to be met by vector $\bm{x}$ to permit valid computation over the MAC. Given these constraints, ChannelComp poses the modulation design method as a feasibility problem to determine the digital modulation vector $\bm{x}$ capable of ensuring a correct computation over-the-air. The resulting digital modulation allows the computation of the desired function using a tabular map.  However, this optimization approach has high complexity when modulation order $q$ is exceedingly high or when the number of nodes $K$ is high~\cite{razavikia2023computing}.

In response to the complexity of the modulation design, we propose a new coding scheme with lower complexity that maintains the full compatibility of ChannelComp with existing digital communication systems. We call such a coding scheme SumComp. It includes digital modulation techniques, such as quadrature-phase shift keying (QPSK), multi-level quadrature amplitude modulation (QAM) (e.g., $4$, $16$, $64$), hexagonal QAM, and pulse-amplitude modulation (PAM). SumComp is an original coding scheme that can be included in the ChannelComp framework to reduce the modulation complexity design optimization problem, which was originally introduced in~\cite{Saeed2023ChannelComp}.

\begin{table}[!t]
\centering
\caption{List of commonly used variables here. Ordered by case and alphabetically.}
\begin{tabular}{|c| c|}
\toprule
\textbf{Variable} & \textbf{Definition} \\
\midrule
$\varphi_k$ & Pre-processing function at node $k$ (source encoder) \\ 
 $\psi$ &   Post-processing function at the CP (source decoder)  \\
 $\mathscr{E}_k$ &   Join source-channel encoder at node $k$  \\
  $\mathscr{D}$ &   Join source-channel decoder   \\
  $\mathcal{G}_{\rho}$ & Modulation mapper \\
$\mathcal{E}_k$ &  Encoder at node $k$ \\
$f$ &  Desired function \\
$\mathcal{D}$  & Decoder at the CP \\
 $K$ & Number of nodes in the network \\
 $q$ & Number of quantization level \\
 $r$ & Received signal by the CP \\  
 $x_k$ & Modulated signal of node $k$\\
 ${z}$ & Additive white Gaussian noise \\
 $\mathcal{X}$ & The set of all constellation points \\
\bottomrule
\end{tabular}
\label{Tab:List}
\end{table}

\subsection{Contributions}

 We propose in this work a coding scheme, which we call SumComp,  based on a ring of integers devised to circumvent the complexities associated with solving the optimization complexity of the modulation design for the ChannelComp over-the-air computation. To diminish the complexity, we focus on computing only the summation function over-the-air, which generalizes to a class of functions known as nomographic functions~\cite{goldenbaum2013reliable} by using pre and post-coders. The SumComp coding uses a ring of integers representing a 2D grid to compute the sum function over the MAC. Through theoretical and numerical results, we demonstrate that our proposed coding scheme can be integrated into various digital modulation schemes. Additionally, we analyze the mean squared error (MSE) associated with using SumComp coding in computing the arithmetic sum function. We also identify an upper limit for the mean absolute error (MAE) within a range of nomographic functions.

Note that while SumComp coding employs a structure that resembles a lattice in its use of a ring of integers representing a conceptual 2D grid, SumComp diverges fundamentally from traditional lattice coding. Traditional lattice codes map data within the same dimensionality (n-D to n-D), maintaining the dimension throughout the process. In contrast, SumComp coding operates as a bandwidth-expanding technique, mapping data from a lower dimension (1D) to a higher one (2D in our case). This expansion emphasizes the significance of the association of grid values rather than the transmitted values. The distinction and the methodological innovation of SumComp coding lie in its unique approach to dimensionality and value association, setting it apart from conventional lattice coding strategies.

Specifically, our contributions are as follows:
\begin{itemize}
    \item \textbf{Digital modulations:} We propose a novel digital modulation coding based on the ring of integers that can compute the sum function for a broad class of digital modulations. SumComp coding exhibits comprehensive compatibility, encompassing various forms of digital modulations. This includes QPSK, QAM at multiple levels (e.g., $4$, $16$, $64$), hexagonal QAM, and other digital modulation techniques.
    \item \textbf{Joint source-channel coding:}  We jointly design the encoding and constellation mapping schemes for the over-the-air computation problem. Specifically, the SumComp coding can be considered a joint source-channel coding for over-the-air computation, where a $1D$ input source signal is mapped into constellation points into a 2D (phase and quadrature components) signal towards computing a function. To the best of our knowledge, SumComp coding is the first joint source-channel coding scheme for digital over-the-air computation.
     \item \textbf{MSE analysis:} We establish the MSE for the SumComp coding in the case of computing the arithmetic sum function. Moreover, we provide an upper bound on the MAE for a class of nomographic functions.   
     \item \textbf{Low complexity coding:}  Different from the findings of \cite{razavikia2023computing}, SumComp coding does not need to solve an optimization problem to obtain the modulation vector, due to its ability at providing a closed-form solution for the encoding procedure.
     \item \textbf{Numerical experiments:}  In addition to the aforementioned advantages, our numerical experiments reveal that SumComp coding, mirroring the performance of ChannelComp, exceeds AirComp in terms of computational accuracy across a range of crucial functions. Notably, in high signal-to-noise ratio (SNR) scenarios, SumComp coding demonstrates a roughly $10$ dB enhancement over other techniques for both sum and product functions. 
\end{itemize}

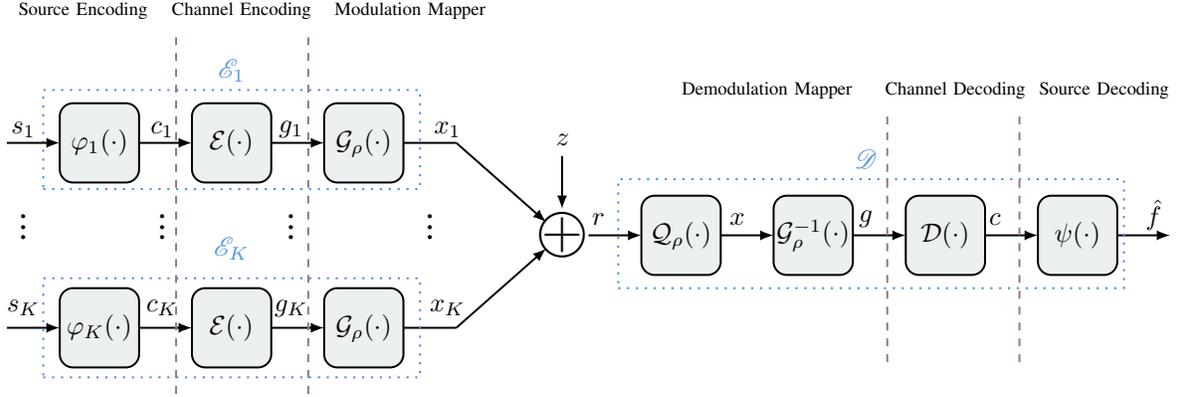
\begin{figure*}
    \centering
\definecolor{palecgray}{rgb}{0.93, 0.94, 0.94}
\tikzset{every picture/.style={line width=0.75pt}} 

\begin{tikzpicture}[x=0.75pt,y=0.75pt,yscale=-1,xscale=1]

\draw  [color={rgb, 255:red, 74; green, 144; blue, 226 }][dash pattern={on 0.84pt off 2.51pt}] (85,80) -- (275,80) -- (275,130) -- (85,130) -- cycle ;

\draw[fill=palecgray , rounded corners=5pt] (470pt, 130pt) rectangle (440pt, 100pt) {};
\draw (455pt,115pt) node   {$\psi(\cdot)$};

\draw [-latex]    (420pt,115pt) -- (440pt,115pt) ;

\draw[fill=palecgray , rounded corners=5pt] (420pt, 130pt) rectangle (390pt, 100pt) {};
\draw (405pt,115pt) node   {$\mathcal{D}(\cdot)$};
\draw [-latex]    (370pt,115pt) -- (390pt,115pt) ;

\draw[fill=palecgray , rounded corners=5pt] (370pt, 130pt) rectangle (340pt, 100pt) {};
\draw (355pt,115pt) node   {$\mathcal{G}_{\rho}^{-1}(\cdot)$};
\draw [-latex]    (320pt,115pt) -- (340pt,115pt) ;

\draw[fill=palecgray , rounded corners=5pt] (320pt, 130pt) rectangle (290pt, 100pt) {};
\draw (305pt,115pt) node   {$\mathcal{Q}_{\rho}(\cdot)$};
\draw [-latex]    (270pt,115pt) -- (290pt,115pt) ;

\draw   (200pt,80pt) -- (220pt,80pt) ;

\draw (260pt, 115pt) node {\LARGE $\bigoplus$};

\draw [-latex]  (470pt,115pt) -- (490pt,115pt) ;

\draw (347,105) node {$z$};

\draw [-latex]  (260pt,85pt) -- (260pt,105pt) ;

\draw [-latex]  (220pt,80pt) -- (254pt,110pt) ;

\draw[fill=palecgray , rounded corners=5pt] (200pt, 95pt) rectangle (170pt, 65pt) {};
\draw (185pt,80pt) node   {$\mathcal{G}_{\rho}(\cdot)$};

\draw [-latex]    (150pt,80pt) -- (170pt,80pt) ;

\draw[fill=palecgray , rounded corners=5pt] (150pt, 95pt) rectangle (120pt, 65pt) {};
\draw (135pt,80pt) node   {$\mathcal{E}(\cdot)$};
\draw [-latex]    (100pt,80pt) -- (120pt,80pt) ;

\draw[fill=palecgray , rounded corners=5pt] (100pt, 95pt) rectangle (70pt, 65pt) {};
\draw (85pt,80pt) node   {$\varphi_1(\cdot)$};
\draw [-latex]    (50pt,80pt) -- (70pt,80pt) ;

\draw    (200pt,150pt) -- (220pt,150pt) ;

\draw[fill=palecgray , rounded corners=5pt] (200pt, 165pt) rectangle (170pt, 135pt) {};
\draw (185pt,150pt) node   {$\mathcal{G}_{\rho}(\cdot)$};

\draw [-latex]    (150pt,150pt) -- (170pt,150pt) ;

\draw[fill=palecgray , rounded corners=5pt] (150pt, 165pt) rectangle (120pt, 135pt) {};
\draw (135pt,150pt) node   {$\mathcal{E}(\cdot)$};
\draw [-latex]    (100pt,150pt) -- (120pt,150pt) ;

\draw[fill=palecgray , rounded corners=5pt] (100pt, 165pt) rectangle (70pt, 135pt) {};
\draw (85pt,150pt) node   {$\varphi_K(\cdot)$};
\draw [-latex]    (50pt,150pt) -- (70pt,150pt) ;

\draw [-latex]  (220pt,150pt) -- (254pt,120pt) ;

\draw  [color={rgb, 255:red, 74; green, 144; blue, 226 }][dash pattern={on 0.84pt off 2.51pt}] (85,175) -- (275,175) -- (275,225) -- (85,225) -- cycle ;

\draw  [color={rgb, 255:red, 74; green, 144; blue, 226 }][dash pattern={on 0.84pt off 2.51pt}] (375,125) -- (630,125) -- (630,180) -- (375,180) -- cycle ;

\draw [dashed, color=bazaar]   (114pt,40pt) -- (114pt,175pt) ;
\draw [dashed, color=bazaar]   (164pt,40pt) -- (164pt,175pt) ;

\draw [dashed, color=bazaar]   (383pt,70pt) -- (383pt,165pt) ;
\draw [dashed, color=bazaar]   (433pt,70pt) -- (433pt,165pt) ;

\draw (180,70) node {\color{rgb, 255:red, 74; green, 144; blue, 226 } $\mathscr{E}_1$};

\draw (105,40) node {\scriptsize Source Encoding};
\draw (185,40) node {\scriptsize Channel Encoding};
\draw (270,40) node {\scriptsize Modulation Mapper};

\draw (75,100) node {$s_1$};
\draw (145,100) node {$c_1$};
\draw (210,100) node {$g_1$};
\draw (289,100) node {$x_1$};

\draw (75,145) node {\Large $\vdots$};
\draw (145,145) node {\Large $\vdots$};
\draw (210,145) node {\Large $\vdots$};
\draw (280,145) node {\Large $\vdots$};

\draw (75,190) node {$s_K$};
\draw (145,190) node {$c_K$};
\draw (210,190) node {$g_K$};
\draw (289,190) node {$x_K$};

\draw (180,160) node {\color{rgb, 255:red, 74; green, 144; blue, 226 } $\mathscr{E}_K$};

\draw (500,115) node {\color{rgb, 255:red, 74; green, 144; blue, 226 } $\mathscr{D}$};

\draw (450,80) node {\scriptsize Demodulation Mapper};
\draw (545,80) node {\scriptsize Channel Decoding};
\draw (620,80) node {\scriptsize Source Decoding};

\draw (365,145) node {$r$};
\draw (435,145) node {$x$};
\draw (500,145) node {$g$};
\draw (565,145) node {$c$};
\draw (645,143) node {$\hat{f}$};

\end{tikzpicture}

    \caption{  Block diagram illustrating the complete transmission process for computing a function over the MAC with $K$ nodes. The process begins with the signals $s_{1}, s_{2}, \ldots, s_{K}$ being passed through an encoder $\mathscr{E}(\cdot)$ and decoder $\mathscr{D}{\rho}$. The modulated signals $x_{1}$, $x_{2}, \ldots x_{K}$ are then transmitted  over the MAC, leading to ${r}$ contaminated by the noise ${z}$. The received signal is ${r}$, which undergoes quantization $\mathcal{Q}(\cdot)$. Then, it is passed through an inverse function $\mathcal{G}_{\rho}^{-1}$ to obtain ${g}\in \mathbb{G}$, which is finally decoded by the decoder $\mathcal{D}(\cdot)$ and processed by function $\psi (\cdot)$ to yield the estimated function $\hat{f}$. The dashed boxes denote the Encoding region $\mathscr{E}$ and the Decoding region $\mathscr{D}$. In a standard communication system, $\varphi$, $\mathcal{E}$, and $\mathcal{G}_{\rho}$ correspond to source coding, channel encoding, and modulation mapper blocks, respectively. Similarly,  $\psi$, $\mathcal{D}$, $\mathcal{G}_{\rho}^{-1}$ corresponds to source decoding, channel decoding, and demodulation mapper blocks, respectively. } 
    \label{fig:Systemmodel}
\end{figure*}

\subsection{Organization of the paper}
The rest of the paper is organized as follows: in Section~\ref{sec:system}, we explain the system model, including the problem statement and the signal model. Next, we describe in detail the architecture of the encoder and decoder of the proposed SumComp coding in Section~\ref{sec:encodedecode}. In Section~\ref{sec:Performance}, we analyze the performance of the SumComp encoder for the sum function and the class of nomographic functions in terms of MSE error and MAE error, respectively, and provide theoretical upper bounds for the  MSE and MAE metrics. We present the numerical results for the theoretical bounds and the performance analysis between SumComp coding, the original coding proposed in ChannelComp, AirComp, and the traditional orthogonal frequency division multiple access (OFDMA) communication in Section~\ref{sec:numerical_results}. Finally, we conclude the paper in Section~\ref{sec:conclusions}.

\subsection{Notation}

We denote a finite field by $\mathbb{F}_q$, where $q$ denotes the number of elements inside the field. Moreover, we denote by $\mathbb{Z}$,  $\mathbb{R}$, and $\mathbb{C}$ as the integer, real, and complex number sets, respectively. Let $\rho$ be a non-zero complex number. We define the ring $\mathbb{Z}[\rho]$ as $\mathbb{Z}[\rho] = \{a+b\rho|~a,b \in \mathbb{Z}\}$.  Then, we denote by $\mathbb{G}$ the Gaussian integer set that is $\mathbb{Z}[i]$, i.e., a Gaussian integer $z = a +bi \in \mathbb{G}$, is a complex number whose real and imaginary parts are integer values, $\{a + bi | a, b \in \mathbb{Z}\}$. Moreover, let $q_1$ and $q_2$ be integers   relatively prime integers or \textit{co-prime}, i.e., their greatest common divisor is ${\rm gcd(q_1,q_2)} = 1$. Then, by Bézout's identity, there exist integers $\mu_1$ and $\mu_2$ such that $q_1\mu_1 + q_2\mu_2  = 1$.

We use lowercase letters $x$ for scalar and calligraphic notation $\mathcal{X}$ to represent operators. For a complex scalar $x\in \mathbb{C}$, $\mathfrak{Re}({x})$ and  $\mathfrak{Im}({x})$ denote the real and imaginary parts of $x$, respectively. Moreover, we use $\mathbb{E} \subset \mathbb{R}$ as the closed unit interval. $\mathcal{F}(\mathbb{A})$ denotes the space of every function $f:\mathbb{A} \mapsto \mathbb{R}$ for some topological space $\mathbb{A}$. Moreover, $\mathcal{C}^{0}(\mathbb{A})$ denotes the space of real-valued continuous functions with domain $\mathbb{A}$.  Furthermore, $\mathcal{CN}(0,\sigma^2)$ denotes a circularly symmetric complex normal distribution, wherein the real and imaginary terms are each distributed according to a normal distribution with variance $\sigma^2$, i.e.,  $\mathcal{N}(0,\sigma^2)$.

\section{System Model}\label{sec:system}

The system model is identical to the one on ChannelComp~\cite{razavikia2023computing}, except that we introduce an improved coding scheme and restrict ourselves to nomographic functions. We study a network composed of $K$ nodes and a computational server, termed the computation point (CP), connected through a common communication channel. This architecture is leveraged to compute a predetermined function $f(s_1,s_2,\ldots,s_K)$ at the CP. Each node $k$ owns a unique input $s_k \in \mathbb{F}_q$, which needs to transmit their $s_k$ values via digital communication. In this system, nodes simultaneously transmit over the MAC to enable the computation of $f$ at the CP.

In this digital transmission scenario, each input $s_k \in \mathbb{F}_q$ undergoes an encoding block through $\mathscr{E}_k(\cdot)$ to produce a digitally modulated signal ${x}_k$, with ${x}_k = \mathscr{E}_k({s}_k)$, where ${x}_k \in \mathbb{C}$ is a complex value whose real and imaginary parts correspond to in-phase and quadrature components. Each node $k$ then transmits ${x}_k$ over the communication channel (see Figure~\ref{fig:Systemmodel}).

Due to the synchronous transmission of all nodes over identical frequencies or codes\footnote{It is assumed that the synchronization among all the nodes and the CP is perfect. Imperfect synchronization can be handled by using the existing techniques,  e.g., \cite{razavikia2022blind,hellstrom2023optimal}.}, the CP server obtains the sum of all ${x}_k$'s through the MAC within a single time slot, as represented in equation~\ref{eq:Aggnoise}, i.e.,
\begin{align}\label{eq:Aggnoise} 
{r} = \sum_{k=1}^{K}h_kp_k{x}_k + {z} \in \mathbb{C}.
\end{align} 
Here, ${r}$ represents the superimposed electromagnetic waves, $h_k$ refers to the channel coefficient between node $k$ and the CP, $p_k$ indicates node $k$'s transmission power, and $z$ is the inherent receiver noise. The receiver noise, $z$, is typically modeled as an additive white Gaussian noise process with zero-mean and variance $\sigma^2$, which is circularly symmetric, i.e., $z \sim \mathcal{CN}(0,\sigma^2)$. Following the power control universally adopted in the over-the-air literature~\cite{cao2020optimal}, we select the transmit power as the inverse of the channel, i.e., $p_{k} = h_{k}^{*}/|h_{k}|^{2}$ for $ k\in [K]$. Hence, Eq.~\eqref{eq:Aggnoise} simplifies to
\begin{align}
    \label{eq:channelfree}
            {r} = \sum_{k=1}^{K}{x}_k+ {z}. 
\end{align}
In the following, we use Eq.~\eqref{eq:channelfree} without losing generality. 

Finally, the estimation of the desired function $f$ can be obtained via a proper decoding scheme $\mathscr{D}$ at the CP, i.e., 
\begin{align}
    \hat{f} := \mathscr{D}(r) = f. 
\end{align}

Notably, this system model parallels the AirComp and ChannelComp~\cite{razavikia2023computing} system models over the MAC, with the only difference being the digital modulation. Thus, this model supports low-latency communication across numerous nodes, as evidenced by~\cite{zhu2019broadband}.  Furthermore,  the encoding and constellation mapping schemes are jointly designed for the over-the-air computation problem.

The goal of this paper is to devise the encoders $\mathscr{E}_k(\cdot)$ and the decoder $\mathscr{D}(\cdot)$ to efficiently compute the desired function $f$ at the CP by harnessing the superposition in \eqref{eq:channelfree}. 
The main challenges are as follows:
\begin{itemize}
  \item In the absence of noise, the received signal ${r}$ by the CP has a reshaped finite constellation diagram, i.e.,  it is a deformation of the original constellation diagram of the transmitting nodes $x_k$ for $k\in [K]$. Then, decoding the desired function $f$ from overlapped constellation signals becomes challenging~\cite{Saeed2023ChannelComp}. 
  \item The superposition of channel input symbols is corrupted by the Gaussian noise. 
  \item  Identifying a set of encoders $\mathscr{E}_k(\cdot)$ and the decoder $\mathscr{D}(\cdot)$ for general function computation, like in ChannelComp, is non-trivial. Thus, we limit the function to the specific class of nomographic functions for closed-form computation. 
  \item Direct design of the encoders $\mathscr{E}_k(\cdot)$ and corresponding decoder is intricate. A modular approach is adopted for separate design (refer to Figure~\ref{fig:Systemmodel}). 
\end{itemize}

In the next section, we propose an architecture for the encoder $\mathscr{E}(\cdot)$ and decoder $\mathscr{D}(\cdot)$ in detail.

\begin{figure*}
    \centering
\definecolor{copperrose}{rgb}{0.05, 0.5, 0.06}
\tikzset{every picture/.style={line width=0.75pt}} 

\begin{tikzpicture}[x=0.65pt,y=0.65pt,yscale=-1]

\draw (0,55) node  {$x_{1}$};

\draw [color={rgb, 255:red, 155; green, 155; blue, 155 } ] [dash pattern={on 4.5pt off 4.5pt}]  (15,70) -- (165,70) ;

\draw[fill=black!90] (100,70) node{}  circle  (3);

\draw[fill=black!90] (70,70) node{}  circle  (3);


\draw[fill=black!90] (40,70) node{}  circle  (3);
\draw[fill=black!90] (130,70) node{}  circle  (3);

\draw (100,55) node  {\tiny $11$};
\draw (40,55) node  {\tiny $00$};
\draw (70,55) node  {\tiny $01$};
\draw (130,55) node  {\tiny $10$};

\draw (40,30) node [color=copperrose]  {$0$};
\draw (70,30) node [color=copperrose ]  {$1$};
\draw (100,30) node [color=copperrose ]  {$3$};
\draw (130,30) node [color=copperrose ]  {$2$};

\draw (40,90)  node {\footnotesize $x_{1}^1$};
\draw (70,90) node {\footnotesize $x_{1}^2$};
\draw (100,90) node {\footnotesize $x_{1}^3$};
\draw (130,90) node {\footnotesize $x_{1}^4$};

\draw (0,160) node  {$x_{2}$};
\draw[-latex]    (180,70) -- (265,97) ;

\draw (270,110) node {\LARGE $\bigoplus$};

\draw [-latex]    (180,175) -- (265,123) ;

\draw[-latex]    (285,110) -- (340,110) ;

\draw [color={rgb, 255:red, 155; green, 155; blue, 155 } ] [dash pattern={on 4.5pt off 4.5pt}]  (15,175) -- (165,175) ;

\draw[fill=black!90] (100,175) node{}  circle  (3);

\draw[fill=black!90] (70,175) node{}  circle  (3);


\draw[fill=black!90] (40,175) node{}  circle  (3);
\draw[fill=black!90] (130,175) node{}  circle  (3);

\draw (100,160) node  {\tiny $11$};
\draw (40,160) node  {\tiny $00$};
\draw (70,160) node  {\tiny $01$};
\draw (130,160) node  {\tiny $10$};

\draw (40,135) node [color=copperrose]  {$0$};
\draw (70,135) node [color=copperrose]  {$1$};
\draw (100,135) node [color=copperrose]  {$3$};
\draw (130,135) node [color=copperrose]  {$2$};

\draw (40,195)  node {\footnotesize $x_{2}^1$};
\draw (70,195) node {\footnotesize $x_{2}^2$};
\draw (100,195) node {\footnotesize $x_{2}^3$};
\draw (130,195) node {\footnotesize $x_{2}^4$};

\draw [color={rgb, 255:red, 155; green, 155; blue, 155 } ] [dash pattern={on 4.5pt off 4.5pt}]  (375,110) -- (650,110) ;


\draw[fill=black!90] (430,110) node{}  circle  (3);
\draw[fill=black!90] (460,110) node{}  circle  (3);
\draw[fill=black!90] (490,110) node{}  circle  (3);
\draw[fill=black!90] (520,110) node{}  circle  (3);
\draw[fill=black!90] (550,110) node{}  circle  (3);
\draw[fill=black!90] (580,110) node{}  circle  (3);
\draw[fill=black!90] (610,110) node{}  circle  (3);

\draw (430,130) node {$r_1$};
\draw (460,130) node {$r_2$};
\draw (490,130) node {$r_3$};
\draw (520,130) node {$r_4$};
\draw (550,130) node {$r_5$};
\draw (580,130) node {$r_6$};
\draw (610,130) node {$r_7$};

\draw (430,85) node [color=copperrose] {$0$};

\draw (460,85) node [color=copperrose] {$1$};

\draw  [color={rgb, 255:red, 208; green, 2; blue, 27 }  ][dash pattern={on 0.84pt off 2.51pt}] (480,50) -- (570,50) -- (570,150) -- (480,150) -- cycle ;
\draw  [color={rgb, 255:red, 155; green, 155; blue, 155 }  ][dash pattern={on 4.5pt off 4.5pt}] (390,160) -- (650,160) -- (650,190) -- (390,190) -- cycle ;

\draw (300,80) node {Over-the-air};


\draw (490,85) node   [color={rgb, 255:red, 208; green, 2; blue, 27 }]  {$3$};
\draw (520,85) node  [color={rgb, 255:red, 208; green, 2; blue, 27 } ]  {$2$};
\draw (610,85) node [color=copperrose  ] {$4$};

\draw (580,85) node  [color=copperrose ] {$5$};
\draw (520,60) node  [color={rgb, 255:red, 208; green, 2; blue, 27 } ]  {$4$};
\draw (550,85) node   [color={rgb, 255:red, 208; green, 2; blue, 27 } ]  {$3$};
\draw (490,60) node [color={rgb, 255:red, 208; green, 2; blue, 27 }  ]  {$2$};
\draw (510,15) node  [color=copperrose  ] {$f(s_{1} ,s_{2}) =s_{1} +s_{2}$};
\draw (550,60) node  [color={rgb, 255:red, 208; green, 2; blue, 27 }   ]  {$6$};

\draw (520,175) node {$r_{3} \ =x_{1}^{1} \ +\ x_{2}^{3} \ =x_{1}^{3} \ +x_{2}^{1} = x_{1}^{2} \ +x_{2}^{2}$};

\draw (480,200) node [color=copperrose ]  {$3$};
\draw (540,200) node [color=copperrose ]  {$3$};

\draw (600,200) node [color=copperrose]  { $2$};

\draw (510,200) node    {$=$};
\draw (570,200) node    {$\neq $};

\end{tikzpicture}


    \caption{  Constellation diagram showcasing the transmission of Gray-coded PAM-modulated signals by two nodes ($K=2$) with $q=4$ ($2$ bits). Destructive overlaps, depicted in red, occur when transmitting values $\{3,2,3\}$ and $\{2,4,6\}$ concurrently, demonstrating the challenges in uniquely decoding the received signals due to interference in the same channel. } 
    \label{fig:PAMGray}
\end{figure*}
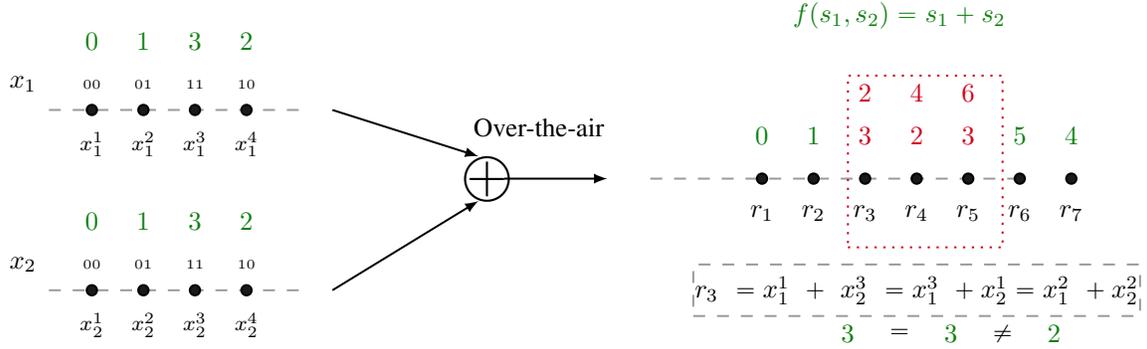

\section{SumComp: Encoding and Decoding Methods}\label{sec:encodedecode}

This section describes how to design the encoder and decoder of our proposed coding scheme, termed SumComp. Before proceeding, we need to briefly recap some necessary definitions and mathematical background provided in the sequel that help the reader understand the results of this paper.

\subsection{Definitions and Motivation}

Here, we target a class of functions called a \textit{nomographic} functions or a set of nomographic functions defined as follows. 
\begin{definition}\label{Def:Nomo}[Nomographic function~\cite{csahin2023survey}]. Let $\mathbb{S}^K$, for $K\geq2$, be a compact metric space. A function  $f : \mathbb{S}^K \mapsto \mathbb{R}$ for which there exist pre-processing functions $\varphi_k: \mathbb{S} \mapsto \mathbb{R}$ for $k\in [K]$ and post-processing function $\psi: \mathbb{R}\mapsto \mathbb{R}$ such that $f$ can be represented by 
    \begin{align}
        f(s_1,\ldots,s_K) = \psi \Big(\sum_{k=1}^{K}\varphi_k(s_k)\Big),
    \end{align}
    is called a nomographic function. 
\end{definition}

Note that the compactness mentioned in Definition~\ref{Def:Nomo} is a critical assumption regarding the commuting of the continuous functions. Indeed, let $\mathcal{N}$, $\mathcal{N}^{0}(\mathbb{E}^{K})$ be the space of nomographic functions and the space of nomographic functions with the restriction of
continuous pre- and post-processing functions, respectively. Then the following interesting result is due to Sprecher and Buck. 
\begin{theorem}\label{th:Sprecher}[Sprecher’65\cite{sprecher1965representation}]. Every function $f\in \mathcal{C}^{0}(\mathbb{E}^{K})$ can be represented with real, monotonic increasing pre-processing functions and possibly a discontinuous post-processing function. 
\end{theorem}
\begin{theorem}\label{th:Buck}[Buck’79\cite{buck1976approximate}]. Every function $f \in \mathcal{F}(\mathbb{E}^{K})$ is nomographic, (i.e., $\mathcal{N}(\mathbb{E}^{K}) = \mathcal{F}(\mathbb{E}^{K}))$. 
\end{theorem}

If one desires the pre- and post-processing functions to be
continuous for an arbitrary continuous function $f$, Theorem~\ref{th:Buck}
is unfortunately not valid. In this case, one may construct a nomographic function approximating the desired function. For example, under Definition~\ref{Def:Nomo}, the geometric mean on $\mathbb{E}^{K}$ is a function that can be approximated with $\varphi_k(s)= \ln{(s+{1}/{p_0(\epsilon)})}$ and $\psi(s)= \exp(s/K)$ for $p_{0}(\epsilon)>0$. For a comprehensive discussion of nomographic functions and their properties, we refer the reader to~\cite{csahin2023survey,goldenbaum2013reliable,sprecher1965structure}.

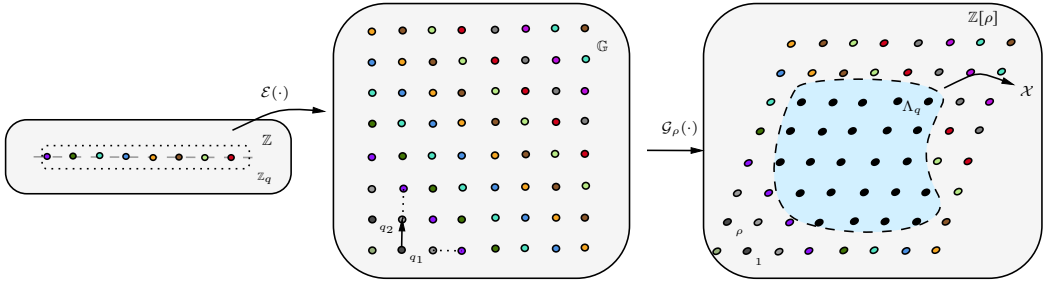
\begin{figure*}
    \centering

\scalebox{0.9}{

\tikzset{every picture/.style={line width=0.75pt}} 

\begin{tikzpicture}[x=0.75pt,y=0.75pt,yscale=-1,xscale=1]

\draw[draw opacity=0][fill={rgb, 255:red, 245; green, 245; blue, 245 } , rounded corners=25pt] (330pt, 135pt) rectangle (490pt, 5pt) {};

\draw[draw opacity=0][fill={rgb, 255:red, 245; green, 245; blue, 245 } , rounded corners=25pt] (155pt, 135pt) rectangle (295pt, 5pt) {};

\draw[draw opacity=0][fill={rgb, 255:red, 245; green, 245; blue, 245 } , rounded corners=10pt] (-60pt, 95pt) rectangle (110pt, 60pt) {};

\draw[draw opacity=1,dash pattern={on 0.84pt off 2.51pt}][rounded corners=4pt] (-28pt, 85pt) rectangle (80pt, 70pt) {};

\draw [color={rgb, 255:red, 155; green, 155; blue, 155 }  ] [dash pattern={on 4.5pt off 4.5pt}]  (-55,103) -- (120,103) ;

\draw  [fill={rgb, 255:red, 74; green, 144; blue, 226 }  ,fill opacity=0.19 ][dash pattern={on 4.5pt off 4.5pt}][line width=0.75]  (499.14,59.34) .. controls (506.57,51.85) and (598.98,51) .. (590.62,69.73) .. controls (582.26,88.47) and (574.12,101.49) .. (587.61,126.87) .. controls (601.09,152.25) and (533.94,156.51) .. (501.45,144.34) .. controls (468.95,132.16) and (491.71,66.84) .. (499.14,59.34) -- cycle ;

\def\spacing{12}
\def\shift{-7}
\def\circleSize{2.2} 


\definecolor{redr}{RGB}{139, 87, 42}
\definecolor{greenr}{RGB}{80, 227, 194}
\definecolor{oranger}{RGB}{245, 166, 35}
\definecolor{bluer}{RGB}{74, 144, 226}
\definecolor{cyanr}{RGB}{184, 233, 134}
\definecolor{jjjdd}{RGB}{208, 2, 27}
\definecolor{tttbbb}{RGB}{65, 117, 8}
\definecolor{grayr}{RGB}{74, 74, 74}
\definecolor{purpler}{RGB}{144, 19, 254}
\definecolor{ddccc}{RGB}{149, 174, 122}
\definecolor{ddyyyy}{RGB}{100, 100, 80}
\definecolor{wwwwyyy}{RGB}{100, 100, 253}

\begin{scope}[shift={(-55,103)}]

\fill[magenta] (0, 0) circle (\circleSize);
\fill[cyan] (\spacing,0) circle (\circleSize);
\fill[jjjdd] (2*\spacing,0) circle (\circleSize);
\fill[tttbbb] (3*\spacing,0) circle (\circleSize);
\fill[grayr] (4*\spacing, 0) circle (\circleSize);
\fill[tttbbb] (5*\spacing, 0) circle (\circleSize);
\fill[greenr] (6*\spacing, 0) circle (\circleSize);
\fill[bluer] (7*\spacing, 0) circle (\circleSize);
\fill[oranger] (8*\spacing, 0) circle (\circleSize);
\fill[redr] (9*\spacing, 0) circle (\circleSize);
\fill[cyanr] (10*\spacing, 0) circle (\circleSize);
\fill[jjjdd] (11*\spacing, 0) circle (\circleSize);
\fill[purpler] (12*\spacing, 0) circle (\circleSize);
\fill[ddccc] (13*\spacing, 0) circle (\circleSize);
\fill[ddyyyy] (14*\spacing, 0) circle (\circleSize);
\fill[wwwwyyy] (15*\spacing, 0) circle (\circleSize);

\end{scope}

\def\spacing{18}

\begin{scope}[shift={(235,30)}]

\foreach \i in {0,...,7} {
          
    \foreach \j in {0,...,7} {
        \ifnum\i=\j
            \fill[oranger] (\i*\spacing, \j*\spacing) circle (\circleSize);
        \fi
        \ifnum\i=\numexpr\j-1
            \fill[bluer] (\i*\spacing, \j*\spacing) circle (\circleSize);
        \fi
        \ifnum\i=\numexpr\j-2
            \fill[greenr] (\i*\spacing, \j*\spacing) circle (\circleSize);
        \fi
        \ifnum\i=\numexpr\j-3
            \fill[tttbbb] (\i*\spacing, \j*\spacing) circle (\circleSize);
        \fi
        \ifnum\i=\numexpr\j-4
            \fill[grayr] (\i*\spacing, \j*\spacing) circle (\circleSize);
        \fi
        \ifnum\i=\numexpr\j-5
            \fill[jjjdd] (\i*\spacing, \j*\spacing) circle (\circleSize);
        \fi
        \ifnum\i=\numexpr\j-6
            \fill[cyan] (\i*\spacing, \j*\spacing) circle (\circleSize);
        \fi
        \ifnum\i=\numexpr\j-7
            \fill[magenta] (\i*\spacing, \j*\spacing) circle (\circleSize);
        \fi
        \ifnum\i=\numexpr\j+1
             \fill[redr] (\i*\spacing, \j*\spacing) circle (\circleSize);
        \fi
        \ifnum\i=\numexpr\j+2
            \fill[cyanr] (\i*\spacing, \j*\spacing) circle (\circleSize);
        \fi
        \ifnum\i=\numexpr\j+3
            \fill[jjjdd] (\i*\spacing, \j*\spacing) circle (\circleSize);
        \fi
        \ifnum\i=\numexpr\j+4
            \fill[purpler] (\i*\spacing, \j*\spacing) circle (\circleSize);
        \fi
        \ifnum\i=\numexpr\j+5
            \fill[ddccc] (\i*\spacing, \j*\spacing) circle (\circleSize);
        \fi
        \ifnum\i=\numexpr\j+6
            \fill[ddyyyy] (\i*\spacing, \j*\spacing) circle (\circleSize);
        \fi
        \ifnum\i=\numexpr\j+7
            \fill[wwwwyyy] (\i*\spacing, \j*\spacing) circle (\circleSize);
        \fi
    }
}

\end{scope}

\begin{scope}[shift={(500,30)}]

\foreach \i in {0,...,7} {
          
    \foreach \j in {0,...,7} {
        \ifnum\i=\j
            \fill[oranger] (\i*\spacing + \j*\shift, \j*\spacing) circle (\circleSize);
        \fi
        \ifnum\i=\numexpr\j-1
            \fill[bluer] (\i*\spacing + \j*\shift, \j*\spacing) circle (\circleSize);
        \fi
        \ifnum\i=\numexpr\j-2
            \fill[greenr] (\i*\spacing + \j*\shift, \j*\spacing) circle (\circleSize);
        \fi
        \ifnum\i=\numexpr\j-3
            \fill[tttbbb] (\i*\spacing + \j*\shift, \j*\spacing) circle (\circleSize);
        \fi
        \ifnum\i=\numexpr\j-4
            \fill[grayr] (\i*\spacing + \j*\shift, \j*\spacing) circle (\circleSize);
        \fi
        \ifnum\i=\numexpr\j-5
            \fill[jjjdd] (\i*\spacing + \j*\shift, \j*\spacing) circle (\circleSize);
        \fi
        \ifnum\i=\numexpr\j-6
            \fill[cyan] (\i*\spacing + \j*\shift, \j*\spacing) circle (\circleSize);
        \fi
        \ifnum\i=\numexpr\j-7
            \fill[magenta] (\i*\spacing + \j*\shift, \j*\spacing) circle (\circleSize);
        \fi
        \ifnum\i=\numexpr\j+1
             \fill[redr] (\i*\spacing + \j*\shift, \j*\spacing) circle (\circleSize);
        \fi
        \ifnum\i=\numexpr\j+2
            \fill[cyanr] (\i*\spacing + \j*\shift, \j*\spacing) circle (\circleSize);
        \fi
        \ifnum\i=\numexpr\j+3
            \fill[jjjdd] (\i*\spacing + \j*\shift, \j*\spacing) circle (\circleSize);
        \fi
        \ifnum\i=\numexpr\j+4
            \fill[purpler] (\i*\spacing + \j*\shift, \j*\spacing) circle (\circleSize);
        \fi
        \ifnum\i=\numexpr\j+5
            \fill[ddccc] (\i*\spacing + \j*\shift, \j*\spacing) circle (\circleSize);
        \fi
        \ifnum\i=\numexpr\j+6
            \fill[ddyyyy] (\i*\spacing + \j*\shift, \j*\spacing) circle (\circleSize);
        \fi
        \ifnum\i=\numexpr\j+7
            \fill[wwwwyyy] (\i*\spacing + \j*\shift, \j*\spacing) circle (\circleSize);
        \fi
    }
}
\end{scope}

\draw[-latex]    (468,157) -- (476,140) ;
\draw[-latex]    (468,157) -- (487,157) ;
\draw[-latex]    (252,156) -- (270,156) ;

\draw[-latex]    (252,156) -- (252,140) ;



\draw (185,80) node   {$\mathcal{E}(\cdot)$};
\draw[-latex]    (165,100) -- (205,100) ;

\draw (637,57) node   [font=\small]  {$\mathcal{X}$};
\draw[-latex]    (590,60) .. controls (610,57) and (605,55) .. (630,57) ;

\draw (140,90) node  [font=\small]  {$\mathbb{Z}$};
\draw (620,15) node [font=\small]  {$\mathbb{Z}[\rho]$};
\draw (375,20) node  [font=\small]  {$\mathbb{G}$};
\draw (420,80) node   {$\mathcal{G}_{\rho}(\cdot)$};
\draw[-latex]    (395,100) -- (435,100) ;

\draw (563.8,66.84) node [anchor=north west][inner sep=0.75pt]  [font=\scriptsize]  {$\Lambda_{q}$};
\draw (252.91,164.21) node [anchor=north west][inner sep=0.75pt]  [font=\tiny]  {$1$};
\draw (110,120) node  [font=\scriptsize]  {$\mathbb{Z}_{q}$};
\draw (240,145) node [anchor=north west][inner sep=0.75pt]  [font=\tiny]  {$1$};
\draw (471.04,165.75) node [anchor=north west][inner sep=0.75pt]  [font=\tiny]  {$1$};
\draw (458.64,147.44) node [anchor=north west][inner sep=0.75pt]  [font=\tiny]  {$\rho$};

\end{tikzpicture}
}

    \caption{The complete encoding procedure. The output of preprocessing function $c_k\in \mathbb{Z}_q$ is first encoded to Gaussian integers $\mathbb{G}$ using the encoder $\mathcal{E}(\cdot)$. Then, using $\mathcal{G}_{\rho}$, we map the output value to a general ring of integers $\mathbb{Z}[\rho]$. Finally, node $k$ selects subset $\Lambda_q$ of $\mathbb{Z}[\rho]$ as the final constellation points to transmit over the MAC. }
    \label{fig:GroupChSum}
\end{figure*}

To compute the nomographic functions, each value $s_k\in \mathbb{E}$ is first encoded to the corresponding pre-processing function $\varphi_k$ to produce the output value $c_k: =\varphi_k(s_k) \in \mathbb{R}$. Because the Gaussian MAC \eqref{eq:channelfree} is a finite capacity channel,  communicating the real  $c_k$ over such a channel with infinite precision is impossible. Hence,  we have to first quantize $c_k$ to finite precision value with $q$ different levels, i.e., $c_k \in \mathbb{F}_q$ for $k\in [K]$. Without loss of generality, we assume the range of $c_k$'s are uniformly quantized so that $\mathbb{F}_q$ can be mapped to $\mathbb{Z}_q$ uniquely using a bijective map~\footnote{Since the domain of $s_k$ is compact, consequently,  the range of preprocessing function $\varphi_k$  is a compact interval. Moreover, the union of these intervals becomes compact as well. As a result, every value in the range of $c_k$'s has a unique dyadic expansion and can be approximated by terminating the dyadic expansion~\cite{goldenbaum2014nomographic}.}. Next,  adding the post-processing function $\psi(\cdot)$ at the end of the decoding scheme, we are able to compute the desired nomographic function $f$ (see Figure~\ref{fig:Systemmodel}).

In the subsequent phase, there is a requirement to map the $c_k \in \mathbb{Z}_q$ onto constellation points $x_k \in \mathbb{Z}^2$ in preparation for transmission. Under typical communication scenarios, Gray coding can map the $q$ potential values of $c_k$ to the constellation points $x_k$, e.g., QAM modulation of order $q$. Subsequently, $x_k$ is transmitted via the MAC to compute the function $f$. However, as demonstrated in~\cite{razavikia2023computing}, the utilization of Gray code in high-order modulation induces destructive interference in the resulting constellation points $\sum_{k=1}^Kx_k$. This ultimately leads to an inability to compute the function $f$ from $r$ using a decoding scheme $\mathscr{D}$, implying  $\hat{f}\neq f$ for any given $\mathscr{D}$.

 To provide a more illustrative example, we present a simple computation scenario for the sum function $f(s_1,s_2) = s_1 + s_2$ for $K=2$ nodes with a Gray-coded PAM $4$ in Figure~\ref{fig:PAMGray}. The constellation points of Gray-coded PAM can be obtained using the following relation:
\begin{align}
    x_k = (s_k+\lfloor s_k/2\rfloor - 2\lfloor s_k/3\rfloor-3 )E_s/2,\quad  k = 1,2,
\end{align}
 where $E_s$ denotes the amplitude of the carrier signal. Then, the constellation points for the $x_1  + x_2$ are given by $r_i = (i-4) \times  E_s$ for $i\in \{1,\ldots,7\}$. Note that the constellation points $r_3$, $r_4$, and $r_5$ need to be assigned simultaneously to different values. Specifically, $r_3$ corresponds to both 3 and 2, $r_4$ corresponds to both $2$ and $4$, and $r_5$ corresponds to both $3$ and $6$.  Therefore, the sum function cannot be computed because the conflict occurs on the abovementioned points.  
 Specifically, in scenarios where nodes $1$ and $2$ transmit respective values of $s_1 = 1$ and $s_2 = 1$ through the MAC, the CP discerns the constellation point $r_3$. Similarly, the transmission of $s_1 = 0$ and $s_2 = 3$ by the nodes also results in the CP perceiving the same constellation point $r_3$. This poses a challenge in the computation of the aggregate $s_1 + s_2$ at the CP, as it necessitates assigning the constellation point $r_3$ to two distinct sums, $3$ and $2$, which is not feasible. Such overlaps prevent precise computation of the desired sum at the CP, showing the inherent limitation of traditional digital AirComp.

In the following Section~\ref{sec:ChSumEncode}, we introduce the working principle of the SumComp encoder and how it can solve destructive overlaps.

\subsection{SumComp Encoder}\label{sec:ChSumEncode}

To encode the constellation points of the modulation $x_k$ to avoid overlaps, we first introduce an encoding scheme that maps any points from $\mathbb{Z}$ to a set of points (line) in a general integer ring $\mathbb{Z}[\rho]$. Then, we show that the induced sets of points belong to a group set under the addition operator. Therefore, the resultant superposition of the constellation points results in a new set of points that belongs to the same set and can be uniquely computed by the summation. Note that the integer set $\mathbb{Z}[\rho]$  helps us to express a general two-dimensional modulation with in-phase and quadrature components.

As outlined in the system model section, the computation of general functions through over-the-air processing poses challenges. A modular approach is employed to circumvent these challenges, specifically targeting nomographic functions. This approach simplifies the computational process by allowing for the independent computation of pre-processing and post-processing functions, thus reducing the requirement to compute the summation function over the air.

Several key factors must be considered regarding the design of an encoder for the summation function. The symmetric nature of the summation function permits the utilization of the same encoder across all nodes, represented as $\mathcal{E}_k(\cdot) = \mathcal{E}(\cdot)$. While it is not necessary, the encoder's ability to operate as an additive map offers considerable benefits, ensuring that $\sum_{k=1}^K\mathcal{E}(s_k) = \mathcal{E}\left(\sum_{k=1}^Ks_k\right)$. Consequently, the objective is to identify a mapping $\mathcal{E}:\mathbb{Z} \mapsto \mathbb{G}$ that enables the output $g_k:= \mathcal{E}(s_k)$, in conjunction with the input value $s_k$, to form an additive group.

We define encoder operator  $ \mathcal{E}_{q_1,q_2} :\mathbb{Z} \mapsto \mathbb{G}$ with fixed positive co-prime integers $(q_1, q_2)$  for given input integer value $m\in \mathbb{Z}$ as  
\begin{align}
\label{eq:Encoder2}
{g}_{k} & :=  \mathcal{E}_{q_1,q_2}(m) =m\mu_{1} + kq_2 +  (m\mu_{2} - kq_1){i} \in \mathbb{G},
\end{align}
for $k\in \mathbb{Z}$, where  $\mu_{1}$ and $\mu_{2}$ are integers such that $1 = q_1 \cdot \mu_{1} + q_2 \cdot \mu_{2}$; they are obtained using the \textit{extended Euclidean algorithm}~\cite{huber1994codes}. The encoder function $\mathcal{E}_{q_1,q_2}(\cdot)$ is defined in terms of  $q_1$ and $q_2$; integers $\mu_{1}$ and $\mu_{2}$ that satisfy Bézout's identity for $q_1$ and $q_2$; and $k$ which varies over all integers. The encoder is an isomorphism map that makes the integers number in $\mathbb{Z}$ and lines in Gaussian integer ring $\mathbb{G}$ isomorphic. 

\begin{rem}

The encoder  $\mathcal{E}_{q_1,q_2}$  serves the purpose of channel coding within communication systems, introducing redundancy by mapping input value points to multiple points (or a line). This added redundancy contributes to the system's resilience against channel noise and facilitates error control, effectively expanding the bandwidth, as elucidated by~\cite{hekland2009shannon}.
\end{rem}
\begin{rem}
      Due to additive map  $\mathcal{E}_{q_1,q_2}$, it is possible to apply the standard channel codes, e.g., non-binary LDPC, and error correction schemes to the SumComp codes to combat the synchronization errors~\cite{xie2023joint}. Furthermore, for the sample-level synchronization, it is possible to design the matched filter to mitigate the error, as studied in~\cite{Shao2022Misaligned,hellstrom2023optimal}.
\end{rem}
Next, for the sake of generality,  we map the encoded value $g_k$ into a general ring of integers $\mathbb{Z}[\rho]$ using an operator $\mathcal{G}_{\rho}:\mathbb{G} \mapsto \mathbb{Z}[\rho]$ for $\rho \in \mathbb{C}$. Indeed, for a given $g = a + bi$, we have
\begin{align}
    {x} = \mathcal{G}_{\rho}(a+bi) = a + b \rho i, \quad a,b \in \mathbb{Z}. 
\end{align}
Then, it is clear that the linear map $\mathcal{G}_{\rho}$ is an isomorphism from $\mathbb{G}$ to the ring of integers $\mathbb{Z}[\rho]$.  Now, we can define the constellation points as a finite subset of generated points in $\mathbb{Z}[\rho]$. Specifically, let $\mathcal{X}$ denote the constellation points generated through the modulation scheme applied to our source messages $c_k$, for $k\in [K]$. To have a finite set of constellation points, we need to restrict the generated lines by encoder $\mathscr{E}_k$ to a given grid subset $\Lambda_q  \subset \mathbb{Z}[\rho]$ with cardinality $|\Lambda_q| \geq q$, for $k\in [K]$.  Mathematically, this relationship is formulated as follows:
\begin{align}
\label{eq:modulation}
\mathcal{X}= \{ \mathcal{G}_{\rho}\big(\mathcal{E}_{q_1,q_2}(c)\big) |~c \in \mathbb{Z}_q\} \cap \Lambda_q. 
\end{align}
This implies that the constellation points can be any arbitrary subset of infinitely generated points by $\mathscr{E}_k$ from $c\in \mathbb{Z}_q$ (see Figure~\ref{fig:GroupChSum}). Consider a point in the domain $\mathbb{Z}_q$, represented as a discrete line in the complex domain $\mathbb{G}$. These corresponding points in $\mathbb{Z}_q$ and $\mathbb{G}$ share the same color in Figure~\ref{fig:GroupChSum}. The mapping $\mathcal{G}_{\rho}$ then translates these discrete lines into the ring of integers, denoted by $\mathbb{Z}[\rho]$. Finally, a finite subset of these points can be selected to constitute a constellation diagram $\mathcal{X}$ (black points in Figure~\ref{fig:GroupChSum}).
 
\begin{rem}
Equation~\eqref{eq:modulation} is sufficiently general to encompass a broad class of digital modulation by selecting an apt grid subset $\Lambda_q$. Two degrees of freedom are evident. Firstly, $\mathcal{G}_{\rho}$ permits flexible grid generation by adjusting $\rho \in \mathbb{C}$. Secondly, the subset $\Lambda_q$ allows for diverse modulation selections through various grid subsets.
\end{rem}
\begin{rem}
It is important to clarify that while $\mathcal{X}$ employs a ring of integers $\mathbb{Z}[\rho]$, SumComp coding significantly differs from traditional lattice coding approaches. Unlike traditional lattice codes, where grid values carry inherent meaning, SumComp emphasizes the association of these values as determined by the encoder $\mathcal{E}_{q_1,q_2}$. Thus, the relationship among grid values, rather than the values themselves, are of primary importance in our scheme. This fundamental shift from value-centric to association-centric encoding~\cite{Saeed2023ChannelComp} distinguishes SumComp from conventional lattice-based methods.
\end{rem}

In what follows, we discuss some examples of the encoder scheme. 

\begin{figure}[!t]
    \centering
    \subfigure[$(q_{1},q_{2}) =(2,3)$]{\label{fig:Hexagonal(a)}
    
\scalebox{0.6}{


\begin{tikzpicture}[x=0.75pt,y=0.75pt,yscale=-1,xscale=1]

\draw[draw opacity=0][fill=pearl , rounded corners=25pt] (100pt, 150pt) rectangle (300pt, 10pt) {};

\draw [color={rgb, 255:red, 155; green, 155; blue, 155 }] [dash pattern={on 4.5pt off 4.5pt}]  (100pt,80pt) -- (300pt,80pt) ;

\draw [color={rgb, 255:red, 155; green, 155; blue, 155 }] [dash pattern={on 4.5pt off 4.5pt}]  (200pt,10pt) -- (200pt,150pt) ;

 \draw[fill=black!90] (110pt,80pt) node{}  circle  (3.5);
 \draw (110pt,70pt) node  {\footnotesize $7$};

\draw[fill=black!90] (170pt,80pt) node{}  circle  (3.5);
\draw (170pt,70pt) node  {\footnotesize $5$};

\draw[fill=black!90] (230pt,80pt) node{}  circle  (3.5);
\draw (230pt,70pt) node  {\footnotesize $3$};

\draw[fill=black!90] (290pt,80pt) node{}  circle  (3.5);
\draw (290pt,70pt) node  {\footnotesize $1$};

\draw[fill=black!90] (170pt,30pt) node{}  circle  (3.5);
\draw (170pt,20pt) node  {\footnotesize $8$};

\draw[fill=black!90] (170pt,130pt) node{}  circle  (3.5);
\draw (170pt,120pt) node  {\footnotesize $2$};

\draw[fill=black!90] (230pt,30pt) node{}  circle  (3.5);
\draw (230pt,20pt) node  {\footnotesize $6$};

\draw[fill=black!90] (230pt,130pt) node{}  circle  (3.5);
\draw (230pt,120pt) node  {\footnotesize $0$};

\draw [color={rgb, 255:red, 74; green, 144; blue, 226 }] (210pt,130pt) node  {\footnotesize $q_1$};

\draw [-latex, color={rgb, 255:red, 74; green, 144; blue, 226 }] (225pt,130pt) .. controls (210pt,145pt) and (190pt,145pt) .. (175pt,130pt) ;

\draw [-latex, color={rgb, 255:red, 74; green, 144; blue, 226 }] (235pt,125pt) .. controls (245pt,110pt) and (245pt,100pt) .. (235pt,85pt) ;

\draw [color={rgb, 255:red, 74; green, 144; blue, 226 }] (250pt,105pt) node  {\footnotesize $q_2$};

\draw  [dash pattern={on 4.5pt off 4.5pt}]  (110pt,130pt) -- (290pt,30pt) ;

\draw[fill=black!30] (110pt,130pt) node{}  circle  (3.5);
\draw (110pt,120pt) node  {\footnotesize $4$};

\draw[fill=black!30] (290pt,30pt) node{}  circle  (3.5);
\draw (290pt,20pt) node  {\footnotesize $4$};

\end{tikzpicture}}
    }\subfigure[$( q_{1} ,q_{2}) \ =\ ( 1,2)$]{\label{fig:Hexagonal(b)}

\tikzset{every picture/.style={line width=0.75pt}} 
\scalebox{0.6}{
\begin{tikzpicture}[x=0.75pt,y=0.75pt,yscale=-1,xscale=1]

\draw[draw opacity=0][fill=pearl , rounded corners=25pt] (100pt, 150pt) rectangle (300pt, 10pt) {};

\draw [color={rgb, 255:red, 155; green, 155; blue, 155 }] [dash pattern={on 4.5pt off 4.5pt}]  (100pt,80pt) -- (300pt,80pt) ;

\draw [color={rgb, 255:red, 155; green, 155; blue, 155 }] [dash pattern={on 4.5pt off 4.5pt}]  (200pt,10pt) -- (200pt,150pt) ;

 \draw[fill=black!90] (110pt,80pt) node{}  circle  (3.5);
 \draw (110pt,70pt) node  {\footnotesize $4$};

\draw[fill=black!90] (170pt,80pt) node{}  circle  (3.5);
\draw (170pt,70pt) node  {\footnotesize $3$};

\draw[fill=black!90] (230pt,80pt) node{}  circle  (3.5);
\draw (230pt,70pt) node  {\footnotesize $2$};

\draw[fill=black!90] (290pt,80pt) node{}  circle  (3.5);
\draw (290pt,70pt) node  {\footnotesize $1$};

\draw[fill=black!90] (170pt,30pt) node{}  circle  (3.5);
\draw (170pt,20pt) node  {\footnotesize $5$};

\draw[fill=black!90] (170pt,130pt) node{}  circle  (3.5);
\draw (170pt,120pt) node  {\footnotesize $1$};

\draw[fill=black!90] (230pt,30pt) node{}  circle  (3.5);
\draw (230pt,20pt) node  {\footnotesize $4$};

\draw[fill=black!90] (230pt,130pt) node{}  circle  (3.5);
\draw (230pt,120pt) node  {\footnotesize $0$};

\draw [color={rgb, 255:red, 74; green, 144; blue, 226 }] (210pt,130pt) node  {\footnotesize $q_1$};

\draw [-latex, color={rgb, 255:red, 74; green, 144; blue, 226 }] (225pt,130pt) .. controls (210pt,145pt) and (190pt,145pt) .. (175pt,130pt) ;

\draw [-latex, color={rgb, 255:red, 74; green, 144; blue, 226 }] (235pt,125pt) .. controls (245pt,110pt) and (245pt,100pt) .. (235pt,85pt) ;

\draw [color={rgb, 255:red, 74; green, 144; blue, 226 }] (250pt,105pt) node  {\footnotesize $q_2$};

\end{tikzpicture}}
    }
    \caption{ SumComp coded Hexagonal QAM of order $8$ for two different choices of $(q_1,q_2)$. In Figure~\ref{fig:Hexagonal(a)}, $(q_1, q_2) = (2, 3)$ shows that the input value $s_k$ can have integer values between $0$ and $8$, except for $4$. Adding the value $4$ requires new constellation points on one of the gray constellation points in the dashed lines, avoiding overlaps. In Figure~\ref{fig:Hexagonal(b)}, the SumComp coded Hexagonal QAM $8$ with $(q_1, q_2) = (1, 2)$ shows that the input value $s_k$ can have integer values between $0$ and $5$. Here, the numbers $4$ and $1$ are repeated. Replacing the numbers $4$ or $1$ with any other numbers results in overlaps among the constellation points.  Note that for the repeated constellation points, every node needs to select one of the constellation points for transmission.}
    \label{fig:Hexagonal}
\end{figure}
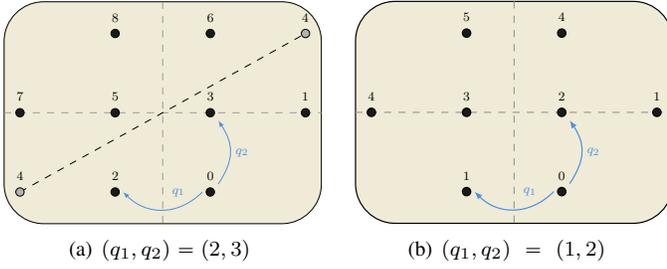

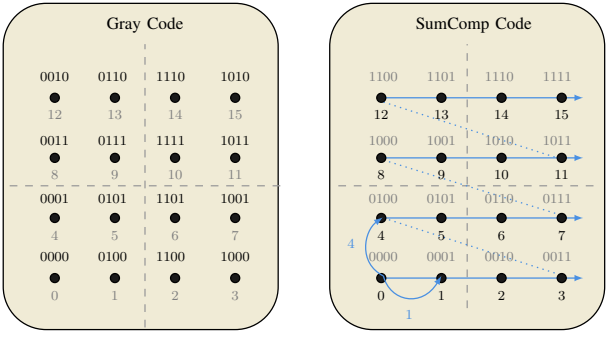
\begin{figure}[t]
    \centering

\definecolor{eggshell}{rgb}{0.94, 0.92, 0.84}
\definecolor{flavescent}{rgb}{0.97, 0.91, 0.56}
\definecolor{lightapricot}{rgb}{0.99, 0.84, 0.69}
\definecolor{peach-orange}{rgb}{1.0, 0.8, 0.6}
\definecolor{peach-yellow}{rgb}{0.98, 0.87, 0.68}
\tikzset{every picture/.style={line width=0.75pt}} 
\scalebox{0.65}{
\begin{tikzpicture}[x=0.75pt,y=0.75pt,yscale=-1]

\draw[draw opacity=0][fill=pearl , rounded corners=25pt] (60pt, 200pt) rectangle (220pt, 10pt) {};

\draw [color={rgb, 255:red, 155; green, 155; blue, 155 } ] [dash pattern={on 4.5pt off 4.5pt}]  (85,155) -- (300,155) ;
\draw [color={rgb, 255:red, 155; green, 155; blue, 155 }] [dash pattern={on 4.5pt off 4.5pt}]  (190,265) -- (190,45) ;

\draw (190,30) node   {Gray Code};

\newcounter{ga}\setcounter{ga}{0}

\foreach \y in {145pt,110pt,75pt,40pt} {

           \foreach \x in {40pt,75pt,110pt,145pt} {
            
          \draw[fill=black!90] (50pt+\x,25pt+\y) node{}  circle  (3.5);
           
           \draw (50pt+\x,35pt+\y) node{\footnotesize \color{black!50} $\thega$} ;

            \stepcounter{ga};
          }

}

\draw (210,120) node  {\footnotesize $1111$};
\draw (260,120) node {\footnotesize $1011$};
\draw (260,70) node   {\footnotesize $1010$};
\draw (210,70) node   {\footnotesize $1110$};
\draw (165,70) node  {\footnotesize $0110$};
\draw (165,120) node {\footnotesize $0111$};
\draw (120,70) node   {\footnotesize $0010$};
\draw (120,120) node  {\footnotesize $0011$};
\draw (165,165) node   {\footnotesize $0101$};
\draw (120,165) node {\footnotesize $0001$};
\draw (120,210) node  {\footnotesize $0000$};
\draw (165,210) node  {\footnotesize $0100$};
\draw (210,210) node   {\footnotesize $1100$};
\draw (260,210) node  {\footnotesize $1000$};
\draw (260,165) node   {\footnotesize $1001$};
\draw (210,165) node  {\footnotesize $1101$};

\draw[draw opacity=0][fill=pearl, rounded corners=25pt] (250pt, 200pt) rectangle (410pt, 10pt) {};

\draw [color={rgb, 255:red, 155; green, 155; blue, 155 } ] [dash pattern={on 4.5pt off 4.5pt}]  (340,155) -- (540,155) ;
\draw [color={rgb, 255:red, 155; green, 155; blue, 155 }] [dash pattern={on 4.5pt off 4.5pt}]  (440,250) -- (440,45) ;

\draw (445,30) node   {SumComp Code};

\draw [color={rgb, 255:red, 74; green, 144; blue, 226 }] [dash pattern={on 0.84pt off 2.51pt}]  (375,90) -- (515,135) ;

\draw [color={rgb, 255:red, 74; green, 144; blue, 226 } ] [dash pattern={on 0.84pt off 2.51pt}]  (375,135) -- (520,180) ;
\draw [color={rgb, 255:red, 74; green, 144; blue, 226 }] [dash pattern={on 0.84pt off 2.51pt}]  (375,180) -- (515,225) ;

\newcounter{ca}\setcounter{ca}{0}

\foreach \y in {145pt,110pt,75pt,40pt} {
            
         \draw[-latex,color={rgb, 255:red, 74; green, 144; blue, 226 }] (375,\y+25pt) -- (530,\y+25pt);

           \foreach \x in {40pt,75pt,110pt,145pt} {
            
          \draw[fill=black!90] (240pt+\x,25pt+\y) node{}  circle  (3.5);
           
           \draw (240pt+\x,35pt+\y) node{\footnotesize $\theca$} ;
           \stepcounter{ca};
          }

}

\draw (465,120) node  {\color{black!50} \footnotesize $1010$};
\draw (510,120) node {\color{black!50}  \footnotesize$1011$};
\draw (510,70) node  {\color{black!50} \footnotesize$1111$};
\draw (465,70) node  {\color{black!50} \footnotesize$1110$};
\draw (420,70) node  {\color{black!50} \footnotesize$1101$};
\draw (420,120) node  {\color{black!50} \footnotesize$1001$};
\draw (375,70) node  {\color{black!50} \footnotesize$1100$};
\draw (375,120) node {\color{black!50} \footnotesize$1000$};
\draw (420,165) node  {\color{black!50} \footnotesize$0101$};
\draw (375,165) node  {\color{black!50} \footnotesize$0100$};
\draw (375,210) node {\color{black!50} \footnotesize$0000$};
\draw (420,210) node  {\color{black!50} \footnotesize$0001$};
\draw (465,210) node  {\color{black!50} \footnotesize$0010$};
\draw (510,210) node  {\color{black!50} \footnotesize$0011$};
\draw (510,165) node  {\color{black!50} \footnotesize$0111$};
\draw (465,165) node  {\color{black!50} \footnotesize$0110$};

\draw (395,255) node  [font=\footnotesize,color={rgb, 255:red, 74; green, 144; blue, 226 } ]  {$1$};
\draw (350,200) node [font=\footnotesize, color={rgb, 255:red, 74; green, 144; blue, 226 }]  {$4$};

\draw [-latex, color={rgb, 255:red, 74; green, 144; blue, 226 }]   (375,225) .. controls (355,215) and (360,190) .. (374.45,179.61) ;

\draw [-latex, color={rgb, 255:red, 74; green, 144; blue, 226 }]   (375,225) .. controls (380,245) and (405,250) .. (420,225) ;

\end{tikzpicture}}

    \caption{Gray code vs SumComp code  for QAM $16$ modulation with $(q_1,q_2) = (1,4)$.}
    \label{fig:GrayCode}
\end{figure}

\begin{exam}
    \label{ex:QAM}
    For QAM with order $q$, which is a perfect square number, the subset $\Lambda_q$ is simply the Cartesian product of $\{1,\ldots \sqrt{q}\}$ with itself. Next, we set $q_2 = \sqrt{q}$ and   $q_1 = 1$ and $\mathcal{G}_{i} = \mathcal{I}$, where  $\mathcal{I}$ is the identity operator. In particular,   the SumComp encoders $\mathscr{E}_k(\cdot)$  become $\mathcal{E}_{\sqrt{q}}:\mathbb{Z}_{q} \mapsto \mathbb{G}$. For input $c\in \mathbb{Z}_q$, we have
    \begin{align}
    \label{eq:EncoderQAM}
   \mathcal{X} & = \{ \mathcal{E}_{1,\sqrt{q}}(c) |~c \in \mathbb{Z}_q\}, \\  \nonumber
   \mathcal{E}_{\sqrt{q}}(c)&:=  c - \sqrt{q} \cdot \Big\lfloor\frac{c}{\sqrt{q}}\Big\rfloor - \sqrt{q}  +  \bigg(\Big\lfloor\frac{c}{\sqrt{q}} \Big\rfloor - \sqrt{q}\bigg)i \in \mathbb{G}.
    \end{align}
The map is bijective. For instance, the encoded values of QAM with order $q = 16$ are depicted in Figure~\ref{fig:GrayCode}.  
\end{exam}

\begin{exam}
    \label{ex:PAM}
    For PAM modulation with order $q$, which is an odd number, the subset $\Lambda_q$ is defined as $\Lambda_q = \{-\lfloor q/2\rfloor, \ldots, 0, \ldots, \lfloor q/2\rfloor \}$ with $q_1 = 1$, $q_2 = q$, and  $\mathcal{G}_{i} = \mathcal{I} $.   The SumComp encoder $\mathcal{E}_{q}(\cdot):\mathbb{Z}_q \mapsto \mathbb{Z}_q$ becomes
    \begin{align}
    \label{eq:EncoderPAM}
    \mathcal{X}  = \{ \mathcal{E}_{1,q}(c) |~c \in \mathbb{Z}_q\},\quad 
   \mathcal{E}_{q}(c):=  c - \lfloor q/2\rfloor, 
    \end{align}
and the decoder $ \mathcal{D}_{q}(x) : \mathcal{X} \mapsto \mathbb{Z}_q$ is defined as
\begin{align}
    \label{eq:decodePAM}
\Tilde{m} := \mathcal{D}_{q}(x):= \mathfrak{Re}(x) + K\lfloor q/2\rfloor, \quad \forall~ {x}\in \mathcal{X}.
\end{align}
\end{exam}

 It is worth mentioning that the subset $\Lambda_{q}$ does not need to be rectangular, and any subset of the generated two-dimensional grid by the encoder $\mathscr{E}(\cdot)$ can be selected. This can be illustrated by  Figure~\ref{fig:Hexagonal}, where we depict the SumComp coded constellation diagram with two different choices of $(q_1, q_2)$ for Hexagonal QAM modulation.  In Figure~\ref{fig:Hexagonal(a)},  the Hexagonal QAM $8$ uses $\mathcal{E}_{2,3}$ for $s_k \in  \{0,\ldots,3,5,\ldots,8\}$. Adding the value $4$ requires new constellation points in the dashed lines on one of the Gray constellation points to avoid overlaps. Therefore, we need to either increase the modulation order or change the choice of $(q_1,q_2)$ in Figure~\ref{fig:Hexagonal(b)}. For the case of $(q_1, q_2) = (1, 2)$, the input domain reduced to the input value $s_k \in \{0,\ldots,5\}$. These examples demonstrate that for a given constellation diagram  (subset $\Lambda_q$), the encoder $\mathscr{E}$ can be a non-surjective mapping. For instance, in Figure~\ref{fig:Hexagonal(b)}, the value $1$ is assigned to two different constellation points. Each node must select only one of the repeated constellation points for transmission to ensure a bijective mapping.

The constellation diagrams presented in Examples \ref{ex:QAM} and \ref{ex:PAM} are not optimized for energy efficiency due to their asymmetry about the origin. However, adjusting the entire constellation diagram with a complex constant can provide symmetry and improve energy efficiency.

\begin{rem}
Note that the encoder's output, $\mathcal{E}(\cdot)$, can undergo a shift by any complex scalar, $\gamma_1 \in \mathbb{C}$, and multiplication with $\gamma_2 \in \mathbb{C}$. Consequently, the decoder, $\mathcal{D}_q$, must adjust the received value by multiplying  $\gamma_2^{*}/|\gamma_2|$ and then subtracting $K\gamma_1$ to maintain the integrity of the computation process, thereby ensuring the successful computation of the summation.
\end{rem}

Note that for computing function $f$ over the MAC, we must show that the constellation points are closed under addition.  To this end, the assigned values to constellation point $x \in \mathcal{X}$ by encoder $\mathscr{E}_k(\cdot)$ form an additive group, which we call the SumComp modulation group. Consequently, additive MAC can not make destructive overlaps as it happens with the Gray code. In particular, we define the SumComp modulation group in Definition~\ref{Df:Modulation}. 
\begin{definition}\label{Df:Modulation}
For a given ring of integers $\mathbb{Z}[\rho]$, the SumComp modulation set $\mathcal{S}_{q_1,q_2}^{\rho}$ is defined  with a precode operator $\mathcal{G}_{\rho}$ and fixed co-prime integers $q_1,q_2\in \mathbb{Z}$ as follows
\begin{align}
\nonumber
    \mathcal{S}_{q_1,q_2}^{\rho} := \Big\{& (c, x) |~x \in \mathbb{Z}[\rho],~c\in \mathbb{Z},\\ &  c  = \mathfrak{Re}(\mathcal{G}_{\rho}^{-1}(x))q_1  +
    \mathfrak{Im}(\mathcal{G}_{\rho}^{-1}(x))q_2 \Big\}, 
\end{align}
where $\mathcal{G}_{\rho}^{-1}$ is an isomorphism that maps points of the ring of integers $\mathbb{Z}[\rho]$ back to the Gaussian integer set~$\mathbb{G}$.  
\end{definition}
In the following, we prove that $\mathcal{S}_{q_1,q_2}^{\rho}$  is a group under the addition. 
\begin{prop}\label{Pr:Group}
The SumComp modulation set $\mathcal{S}_{q_1,q_2}^{\rho}$ is an additive group.
\end{prop}
\begin{proof}
    See Appendix~\ref{Ap:GroupProof}. 
\end{proof}

Note that $\mathcal{S}_{q_1,q_2}^{\rho}$ is a group whose elements are the values of the input signal $c$ with its corresponding constellation points $x$. Proposition~\ref{Pr:Group}  proves that $\mathcal{S}_{q_1,q_2}^{\rho}$ is a group under addition, which means that the summation $\sum_{k=1}^Kc_k$ with corresponding pair $\sum_{k=1}^Kx_k$ also belongs to $\mathcal{S}_{q_1,q_2}^{\rho}$. As a result, using a proper isomorphic decoding scheme $\mathscr{D}$, we assign $\sum_kx_k$ to  $f = \sum_ks_k$ and compute the desired summation function.

The next section explains the overall decoding procedure to compute the function~$f$.

\vspace{20pt}
\subsection{SumComp Decoder}\label{sec:ChSumDecode}

Let $\mathcal{X}^{(K)}$ denote the finite set of
all feasible noiseless aggregate constellation points, i.e.,
\begin{align*}
    \mathcal{X}^{(K)}
:=
\left\{
\sum_{k=1}^{K}x_k \;\middle|\; x_k\in\mathcal{X},\ k\in[K]
\right\}. 
\end{align*}
After transmitting $x_k$'s by nodes, the noiseless aggregate point $r$ in $\mathbb{Z}[\rho] \cap \mathcal{X}^{(K)}$ deviates due to channel noise and no longer belongs to the same integer grid. To map $r \in \mathbb{C}$ back to the grid $\mathbb{Z}[\rho]$, we  use the associated \emph{quantizer} operator to the ring of integers $\mathbb{Z}[\rho]$ denoted by $\mathcal{Q}_{\rho} : \mathbb{C} \mapsto \mathbb{Z}[\rho]$ and defined as
\begin{align}
    \label{eq:QuntizerRing}
    \mathcal{Q}_{\rho}(\mu) = \underset{x \in \mathbb{Z}[\rho]\cap \mathcal{X}^K}{\rm argmin}\|\mu-x\|_2,
\end{align}
where $\mathcal{Q}_{\rho}$ assigns every $\mu \in \mathbb{C}$ to the nearest point, with respect to the Euclidean distance, the ring of integers $\mathbb{Z}[\rho]$. 
Hence, using $\mathcal{Q}_{\rho}$, we obtain
$x = \mathcal{Q}_{\rho}(r)\in \mathbb{Z}[\rho]$. However, in the SumComp
decoder, not every point in $\mathbb{Z}[\rho]$ is a valid decoding point.
Since each transmitter selects its symbol from a finite constellation subset,
the noiseless aggregate received point belongs to a finite aggregate SumComp
set. Therefore, the decoder is defined only over the valid finite subset of
$\mathcal{S}_{q_1,q_2}^{\rho}$ induced by the $K$-fold sum of the transmitted
constellation points. 

Then, the recovered point is interpreted as a point
$x\in \mathcal{X}^{(K)}$ associated with an element of
$\mathcal{S}_{q_1,q_2}^{\rho}$. The decoder
$\mathcal{D}_{q_1,q_2}$ then computes the weighted sum of the real and
imaginary components after mapping $x$ back to the Gaussian integer domain
through $\mathcal{G}_{\rho}^{-1}$. Specifically,
\begin{align}
    \nonumber
\hat{c} & := \mathcal{D}_{q_1,q_2}(\mathcal{G}_{\rho}^{-1}(x)), \\ \label{eq:decode2}
& :=  \mathfrak{Re}(\mathcal{G}_{\rho}^{-1}(x))q_1 + \mathfrak{Im}(\mathcal{G}_{\rho}^{-1}(x))q_2,\quad q_1,q_2 \in \mathbb{Z}
\end{align}
where $x = \mathcal{Q}_{\rho}(\mu)$ is the output of the quantizer $\mathcal{Q}_{\rho}$. Therefore, by defining the decoding block as operator $\mathscr{D}: =\psi\mathcal{D}_{q_1,q_2}\mathcal{G}_{\rho}^{-1}\mathcal{Q}_{\rho}$ that acts on the received signal at the receiver, the overall decoding procedure can be expressed as follows:
\begin{align}
   \hat{f} = \mathscr{D}(r) :=  \psi\Big(\mathcal{D}_{q_1,q_2}\big(\mathcal{G}_{\rho}^{-1}(\mathcal{Q}_{\rho}(r)\big)\Big).
\end{align}
Next, we verify that the pair of $(\mathcal{E},\mathcal{D})$ accurately reproduces the original input. For the noiseless scenario, i.e., $r = \sum_{k=1}^Kx_k$, we can check that $f  = \mathscr{D}(\sum_{k=1}^Kx_k)$. Specifically,  we can write
\begin{align}
\nonumber
 \mathscr{D}\Big(\sum\nolimits_{k=1}^Kx_k\Big) & = \psi\bigg(\mathcal{D}_{q_1,q_2}\Big(\mathcal{G}_{\rho}^{-1}\big(\sum\nolimits_{k=1}^Kx_k\big)\Big)\bigg), \\ \nonumber
        & =  \psi\bigg(\mathcal{D}_{q_1,q_2}\Big(\sum\nolimits_{k=1}^K\mathcal{G}_{\rho}^{-1}(x_k)\Big)\bigg), \\ \nonumber & = \psi\bigg(\mathfrak{Re}\Big(\sum_{k=1}^Kg_k\Big)q_1 + \mathfrak{Im}\Big(\sum_{k=1}^Kg_k\Big)q_2\bigg), \\ \nonumber
        & =\psi\bigg( \sum\nolimits_{k=1}^K\Big(\mathfrak{Re}(g_k)q_1 + \mathfrak{Im}(g_k) q_2\Big)\bigg), \\  & =  \psi\Big(\sum_{k=1}^Kc_k\Big) = \psi\Big(\sum_{k=1}^K \varphi_k(s_k)\Big) =f,
\end{align}
where the first equality comes from the fact that without noise, the quantizer operator  $\mathcal{Q}$ acts as an identity operator. Moreover, the second equality is because of the linearity of the operator $\mathcal{G}_{\rho}$. In the presence of noise, the decoded function obviously results in error and $\hat{f} \neq f$. In the next section, we analyze the impact of the noise on the SumComp coding in terms of mean square and absolute error for different functions with various hyperparameters as $q_1$, $q_2$, and $\rho$. The overall encoding and decoding procedure for SumComp coding is presented in Algorithm~\ref{alg:SumComp}.

\begin{algorithm}[!t]
\caption{SumComp coding: digital coding over-the-air}\label{alg:SumComp}
\begin{algorithmic}[1]
	\State \textbf{Input:} Function $f(s_1,\ldots,s_K)$, $\rho$, modulation order $q$, $q_1$ and $q_2$  encoding scheme  
	\State \textbf{Output:} output of function $\hat{f}$
        \Initialize{\strut {CP shared $\varphi_k$  functions, $q_1,q_2$ and $\rho$ with the nodes \\
        Node $k$ computes $\mathcal{E}$ and $\mathcal{G}_{\rho}$ by $q_1,q_2$ and $\rho$}}
        
	\Procedure{SumComp}{$s_1,\ldots,s_K$}
	\For {{in parallel} $k \gets 1,2,\ldots,K$}
	\State Node $k$ computes $c_k = \varphi_k(s_k)$
        \State Node $k$ computes constellation point  $x_k$ using \eqref{eq:modulation}
        \State Node $k$ transmit $x_k$ simultaneously for OAC
	\EndFor
         \If{CP received $r$ from \eqref{eq:Aggnoise}}
	\State  CP maps $r$ to the nearest neighbor in $\mathbb{Z}[\rho]$ to obtain $\hat{x}$ using \eqref{eq:QuntizerRing}. 
	\State CP computes $\hat{c}_k$ using \eqref{eq:decode2}
        \State CP obtains output by $\hat{f} = \psi\big(\sum_k\hat{c}_k\big)$
	\EndIf
	\EndProcedure
\end{algorithmic}
\end{algorithm}

\section{SumComp Coding: Performance analysis}\label{sec:Performance}

In this section, we analyze the SumComp coding in terms  of
desired function recovery from $\hat{f}$.  It is clear that the analysis depends strongly on the desired function and, thus, on the post-processing function $\psi$, which is generally nonlinear. A unified error analysis is, therefore, very complicated and has to be done separately for every individual $f$. Consequently, we first focus on the simple, but important, special case of the \textit{arithmetic sum} with helpful results on the error. Then, we extend the results to a larger class of nomographic functions where the post-processing function $\psi$ is a uniformly continuous function. 
\subsection{MSE and MAE Analysis}

To analyze the computation error, we use the classical MSE and MAE, which are respectively defined as 
\begin{align}
    {\rm MSE}(\hat{{f}}) & := \mathbb{E}\big\{|f - \hat{f}|^2\big\},\\
    {\rm MAE}(\hat{{f}}) & := \mathbb{E}\big\{|f - \hat{f}|\big\},
\end{align}
where the expectation is calculated over the randomness of the noise, $\hat{f} = \mathcal{D}(r)$ is the estimated value of the summation function at the receiver, and $f$ is the true value, i.e., $f = \psi \Big(\sum_{k=1}^{K}\varphi_k(s_k)\Big)$.   For the case where the desired function is the arithmetic sum case, or equivalently,  $\varphi(s_k) =s_k$ for $k\in [K]$ and $\psi(g) = g $. Then, we have the following result regarding the MSE. 

\begin{prop}
    \label{Pr:MSE}
    Consider a communication network with $K$ nodes where the nodes use the SumComp encoder $\mathscr{E}$ to compute the summation $f = \sum_{k}s_k$, where $s_k\in \mathbb{Z}_q$ over the noisy MAC with $\mathcal{CN}(0,\sigma^2)$. Assuming the noiseless aggregate constellation point is uniformly
distributed over the rectangular aggregate grid $\mathcal{X}^{(K)}$,   the MSE of the computation errors of using the constellation points set $\mathcal{X}^{(K)}$ with uniform rectangular $\Lambda_q = [M_1] \times [M_2]$ for two positive integers $M_1, M_2 \in \mathbb{Z}^{+}$, is given by 
    \begin{align}
     \label{eq:MSEbound}
        {\rm MSE}(\hat{f})  = q_1^2\alpha_{1} + q_2^2\alpha_{2},
    \end{align}
    where
\begin{align}
\alpha_{1}
&=
2\sum_{\ell_1=1}^{M_1-1}
(2\ell_1-1)
\Big(1-\frac{\ell_1}{M_1}\Big)
Q\Big(\frac{2\ell_1-1}{2\sigma}\Big),\\
\alpha_{2}
&=
2\sum_{\ell_2=1}^{M_2-1}
(2\ell_2-1)
\Big(1-\frac{\ell_2}{M_2}\Big)
Q\Big(\frac{(2\ell_2-1)|\rho|}{2\sigma}\Big).
\end{align}
    where $q_1,q_2\in\mathbb{Z}^{+}$ are coprime integers, and
$\rho\in\mathbb{R}^{+}$ determines the spacing of the second coordinate of
the rectangular ring grid.  Moreover, $Q(x)$ is the Gaussian $Q$ function, i.e., 
\begin{align}
    Q(x)  = \frac{1}{\sqrt{2\pi}}\int_{x}^{\infty}{\rm e}^{-\frac{t^2}{2}}dt.
\end{align}

\end{prop}
\begin{proof}
    See Appendix~\ref{Ap:MSEProof}.
\end{proof}

Proposition \ref{Pr:MSE} delineates that MSE comprises two components, each influenced by the selection of $q_1$ or $q_2$. Modifying $q_1$ or $q_2$ inversely affects the counterpart and alters one of the error components. Additionally, a reduction in $q_1$ or $q_2$ elevates $\alpha_1$ or $\alpha_2$, correspondingly. The optimal selection of $q_1$ or $q_2$, contingent upon the chosen $\rho$, can achieve minimal MSE.

\begin{rem}
Assuming a uniform distribution across the constellation points induced by $\sum_{k}s_k$, the MSE specified in \eqref{eq:MSEbound} represents the expected error value. Furthermore, when $\sigma/Kq \ll 1$, where $K$ denotes the number of nodes and $q$ the modulation order, the MSE provided in \eqref{eq:MSEbound} holds, regardless of the prior distribution of $s_k$ or the specific constellation points.
\end{rem}

Then, the obtained analytical expression of the MSE can be extended to a class of \textit{uniformly continuous} functions~\cite{aronszajn1956extension} defined as follows.
\begin{definition}
A function $\psi: \mathbb{S} \mapsto \mathbb{R}$ is  called uniformly continuous function if there exists a strictly increasing concave function $w_{\psi}: [0,\infty) \mapsto [0,\infty)$ with $w_{\psi}(0) = 0$ such that, for all real $x, y\in \mathbb{S}$, we have
\begin{align}
    |\psi(x)-\psi(y)|\leq  w_{\psi}(|x-y|),
\end{align}
where $w_{\psi}$ is called the modulus of continuity of $\psi$. 
\end{definition}
Some examples of these functions are listed below:
\begin{itemize}
    \item For $L$-Lipschitz function $\psi: \mathbb{R} \mapsto \mathbb{R}$, the modulus function becomes $w_{\psi}(x)= Lx$.
    \item If $\psi$ is a Hölder continuous with $(C,\alpha)$ for $\alpha \in (0,1]$, i.e., for all $x, y$ in the domain of $\psi$, we have $|\psi(x)-\psi(y)|\leq C|x-y|^{\alpha}$, then  $w_{\psi}$ reads to $x \mapsto Cx^{\alpha}$. 
    \item When $\psi$ is an increasing concave function, then $w_{\psi}$ becomes trivial, i.e., $w_{\psi}: x \mapsto x$. 
\end{itemize}
In the following, we have an upper bound on the MAE for the nomographic function with uniformly continuous post-processing function $\psi$. 
\begin{prop}
    \label{Pr:MAE}
    Consider a communication network with $K$ nodes, where node $k$ uses the encoder $\mathscr{E}_{k}$ to compute the summation $f = \psi \Big(\sum_{k=1}^{K}\varphi_k(s_k)\Big)$,  and $\psi$ is an uniformly continuous function over the noisy MAC with $\mathcal{CN}(0,\sigma^2)$. Also, assume  the noiseless aggregate constellation point is uniformly distributed over the rectangular aggregate grid $\mathcal{X}^{(K)}$. Then, the MAE of the computation errors when utilizing the constellation points from the set $\mathcal{X}$ with $\Lambda_q$, as in Proposition~\ref{Pr:MSE}, is confined within the following upper bound
    \begin{align}
        {\rm MAE}(\hat{f}) \leq w_{\psi}(q_1\beta_{1} + q_2\beta_{2}),
    \end{align}
    where $w_{\psi}$ signifies  the modulus of continuity of $\psi$, and $\beta_{1}$ and $\beta_{2}$ are given by
    \begin{subequations}
    \label{eq:Beta1Beta2}
\begin{align}
    \beta_1 & = 2 \sum\nolimits_{\ell_1=1}^{M_1-1}\Big(1 -\frac{\ell_1}{M_1}\Big) Q\Big(\frac{(2\ell_1-1)}{2\sigma}\Big), \\
   \beta_2 & =  2 \sum\nolimits_{\ell_2=1}^{M_2-1}\Big(1 - \frac{\ell_2}{M_2}\Big) Q\Big(\frac{(2\ell_2-1)|\rho|}{2\sigma}\Big).
   \end{align}
\end{subequations}
\end{prop}
\begin{proof}
  See Appendix~\ref{Ap:MAEProof}.
\end{proof}

Different from Proposition~\ref{Pr:MSE}, Proposition~\ref{Pr:MAE} establishes that the MAE of $\hat{f}$ is subject to an upper limit rather than an average value. Nevertheless, this upper bound elucidates a compromise between the selections of $q_1$ and $q_2$. The relationship of this upper bound with $q_1$ and $q_2$ varies according to the chosen $w_{\psi}$ function, potentially exhibiting linear, quadratic, or other forms of dependency. This differs from the quadratic relationship identified in the MSE scenario.

\begin{cor}
Consider that $\psi$ is Hölder continuous with $(C,\alpha)$. Then, for any nomographic function $f$ with post-processing $\psi(\cdot)$, the MAE is upper bounded by
\begin{align}
{\rm MAE}(\hat{f}) \leq C |q_1\beta_{1} + q_2\beta_{2}|^{\alpha},
\end{align}
where $\alpha \in (0,1]$. 
\end{cor}
To illustrate, we take a few examples here.
\begin{exam}(Maximum) The maximum function $f = \max_ks_k$ can be approximated with  $\varphi_k(s_k) = \exp{(s_k)}$ and  $\psi(x) = \ln{(x)}$. Then,  $\psi(x)$ is a concave and strictly increasing function. Therefore, $w_{\psi}$ is the identity function. However, if we bound the min of $x$ by a factor $\theta$, i.e., $\theta \leq x$, then the logarithmic function becomes Lipschitz continuous with ${1}/{\theta}$. Therefore, we have
\begin{align}
{\rm MAE}(\hat{f}) \leq  \frac{q_1\beta_{1} + q_2\beta_{2}}{\theta}.
\end{align}
\end{exam}
\begin{exam}\label{ex:Euclidean}(Euclidean norm) Let the desired function be the
 Euclidean norm, i.e.,  $f = \sqrt{s_1^2 + \ldots + s_K^2}$ where $\varphi_k(s_k) =s_k^2$ and $\psi(x) = \sqrt{x}$. Hence, $\psi$ is Hölder continuous with $\alpha = 0.5$, and consequently, it yields
\begin{align}
{\rm MAE}(\hat{f}) \leq  \sqrt{|q_1\beta_{1} + q_2\beta_{2}|}.
\end{align}
\end{exam}
\begin{exam}(Arithmetic mean)\label{ex:Arithmetic}
Let function f be the arithmetic mean function, i.e., $\sum_{k=1}^Ks_k/K$. Then $\varphi_k(s_k) = s_k$ and  $\psi(x) = {x}/{K}$ which is a concave and strictly increasing function.  As a result, $w_{\psi}$ is ${1}/{K}$, i.e.,
\begin{align}
{\rm MAE}(\hat{f}) \leq  \frac{q_1\beta_{1} + q_2\beta_{2}}{K}.
\end{align}
\end{exam}

\subsection{Complexity Analysis}

In this section, we compare the complexity of SumComp coding to the coding scheme proposed in  ChannelComp in terms of the number of basic operators (BOPs) to show the obtained reduction in the cost. Indeed, ChannelComp~\cite{Saeed2023ChannelComp} uses a coding that is based on an optimization problem where the input is a vector involving the constellation points of the modulation diagram of all the nodes to obtain the encoder $\mathscr{E}_k$ for all the nodes. Also, the decoder $\mathscr{D}$ is then determined as a tabular function that maps the resultant constellations to the output co-domain of the function $f$. 

In the sequel, we analyze the complexity of the encoder and decoder. Consider that the input vector is of size $q\times K$, where $q$ denotes the quantization levels and  $K$ is the number of nodes in the network. The complexity of solving a semidefinite programming optimization in ChannelComp is $\mathcal{O}(\max\{n,m\}^4n^{0.5})$~\cite{ye2011interior,luo2010semidefinite}, in which $n$ is the dimension of the input variable, i.e., $q\times K$, and $m$ is the number of constraints, which is at most $ m \leq q^K(1+q^K)/2$. As a result, the computational complexity to find a valid modulation vector in ChannelComp is at most $\mathcal{O}(q^{0.5}K^{0.5}q^{8K})$, or simply $\mathcal{O}(\sqrt{K}q^{8K+0.5})$, for computing a general function. Moreover, for the case of symmetric functions,  the modulation vectors are the same for all $K$ nodes,  which implies that $n = q$.  Further, the number of constraints becomes at most $m\leq \binom{K+q-1}{q-1} < \exp{(K+q-1)}$~\cite{Saeed2023ChannelComp}. Therefore, the overall computational complexity reduces to $\mathcal{O}(\sqrt{q}{\rm e}^{4(K+q-1)})$.

For analyzing the complexity of the decoder part, we need to know the number of constellation points, which depends on the output of the optimization problem and can not be determined beforehand. Hence, it is difficult to consider an exact number. However, the number of constellation points can reach at most $q^K$ different values. In this case, using a linear search for the tabular function, ChannelComp can decode the received signal at most in $\mathcal{O}(q^K)$  BOPs~\cite{knuth1973sorting}.

\begin{figure*}[!t]
\centering
\subfigure[QAM modulation]{
    \label{fig:SUMQAM}
    \begin{tikzpicture} 
    \begin{axis}[
        xlabel={${\rm SNR (dB)}$},
        ylabel={${\rm MSE}(\hat{f})$},
        label style={font=\scriptsize},
        tick label style={font=\scriptsize} , 
        width=0.49\textwidth,
        height=4.5cm,
        xmin=-15, xmax=10,
        ymin=5e-5, ymax=10^3,
        xtick={-10, -5, 0, 5, 10,15},
         ymode = log,
       legend style={nodes={scale=0.55, transform shape}, at={(0.99,0.97)}}, 
        ymajorgrids=true,
        xmajorgrids=true,
        grid style=dashed,
        grid=both,
        grid style={line width=.1pt, draw=gray!15},
        major grid style={line width=.2pt,draw=gray!40},
    ]
     \addplot[only marks, 
        color=bazaar,
        mark=triangle,
        mark options = {rotate = 180},
        line width=1pt,
        mark size=2pt,
        ]
    table[x=SNR,y=MSES1]
    {Data/MSEError.dat};
    \addplot[
        color=bazaar,
        mark=triangle,
        line width=1pt,
        mark size=2pt,
        ]
    table[x=SNR,y=MSET1]
    {Data/MSEError.dat};
     \addplot[only marks, 
        color=burntsienna,
        mark=pentagon,
        mark options = {rotate = 36},
        line width=1pt,
        mark size=2pt,
        ]
    table[x=SNR,y=MSES2]
    {Data/MSEError.dat};
    \addplot[
        color=burntsienna,
        mark=pentagon,
        line width=1pt,
        mark size=2pt,
        ]
    table[x=SNR,y=MSET2]
    {Data/MSEError.dat};
     \addplot[ only marks,
        color=bulgarianrose,
        mark=o,
        line width=1pt,
        mark size=2pt,
        ]
    table[x=SNR,y=MSES3]
    {Data/MSEError.dat};
    \addplot[
        color=bulgarianrose,
        mark=star,
        line width=1pt,
        mark size=2pt,
        ]
    table[x=SNR,y=MSET3]
    {Data/MSEError.dat};
    \legend{QAM $16$ (empirical),  QAM $16$ (analytical), QAM $64$ (empirical), QAM $64$ (analytical), QAM $256$ (empirical), QAM $256$ (analytical)};
    \end{axis}
\end{tikzpicture}}\subfigure[PAM modulation]{
    \label{fig:SUMPAM}
    \begin{tikzpicture} 
    \begin{axis}[
        xlabel={${\rm SNR (dB)}$},
        ylabel={${\rm MSE}(\hat{f})$},
        label style={font=\scriptsize},
        tick label style={font=\scriptsize} , 
        width=0.49\textwidth,
        height=4.5cm,
        xmin=-15, xmax=19,
        ymin=7e-3, ymax=10^4,
        xtick={-10, -5, 0, 5, 10,15},
         ymode = log,
       legend style={nodes={scale=0.55, transform shape}, at={(0.99,0.97)}}, 
        ymajorgrids=true,
        xmajorgrids=true,
        grid style=dashed,
        grid=both,
        grid style={line width=.1pt, draw=gray!15},
        major grid style={line width=.2pt,draw=gray!40},
    ]
     \addplot[ only marks,
        color=bluebell,
        mark=+,
        mark options = {rotate = 180},
        line width=0.75pt,
        mark size=2pt,
        ]
    table[x=SNR,y=MSES4]
    {Data/MSEError.dat};
    \addplot[
        color=bluebell,
        mark=x,
        line width=0.75pt,
        mark size=2pt,
        ]
    table[x=SNR,y=MSET4]
    {Data/MSEError.dat};
     \addplot[ only marks,
        color=darklavender,
        mark=triangle,
        mark options = {rotate = 180},
        line width=1pt,
        mark size=2pt,
        ]
    table[x=SNR,y=MSES5]
    {Data/MSEError.dat};
    \addplot[
        color=darklavender,
        mark=triangle,
        line width=1pt,
        mark size=2pt,
        ]
    table[x=SNR,y=MSET5]
    {Data/MSEError.dat};
     \addplot[only marks,
        color=darkcerulean,
        mark=square,
        mark options = {rotate = 45},
        line width=1pt,
        mark size=2pt,
        ]
    table[x=SNR,y=MSES6]
    {Data/MSEError.dat};
    \addplot[
        color=darkcerulean,
        mark=square,
        line width=1pt,
        mark size=2pt,
        ]
    table[x=SNR,y=MSET6]
    {Data/MSEError.dat};
    \legend{PAM $16$ (empirical), PAM $16$  (analytical), PAM $32$ (empirical), PAM $32$ (analytical), PAM $64$ (empirical), PAM $64$ (analytical)};
    \end{axis}
\end{tikzpicture}

}
  \caption{Monte Carlo numerical evaluation of the MSE of computing the summation function for $5 \times 10^4$ trials versus the analytical results from Proposition \ref{Pr:MSE} for digital modulation. Figures \ref{fig:SUMQAM} and \ref{fig:SUMPAM} show both empirical and analytical MSE for QAM modulations of order $q = \{16, 64, 256\}$, and PAM modulation of order  $q = \{16, 32, 64\}$, respectively.}
  \label{fig:SumSim}
\end{figure*}
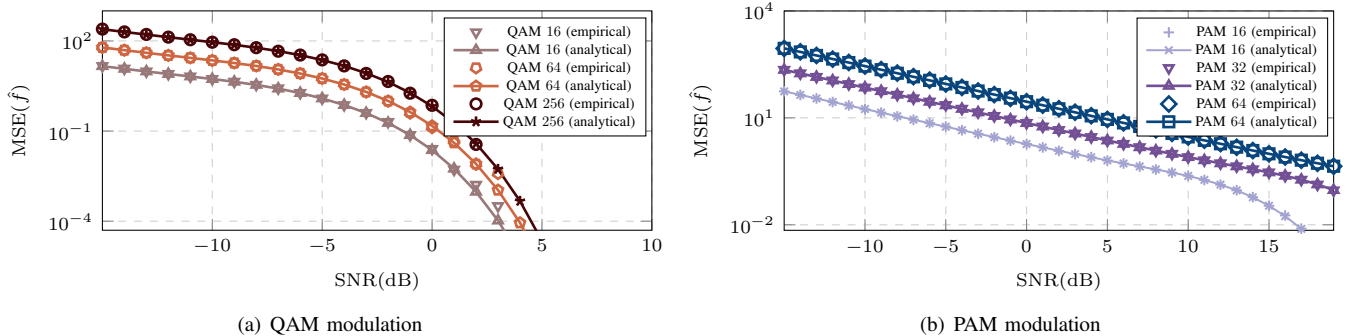

However, SumComp coding offers an optimization-free approach. To analyze the complexity of the encoder $\mathscr{E}_k$ including the composite operator $\mathcal{G}_{\rho}\mathcal{E}\varphi_k$, we need to analyze each operator separately. Similarly, we should consider a similar approach for the decoder $\mathscr{D}$ with its sub-operators. The following proposition establishes the complexity of a class of nomographic functions.

\begin{prop}
    \label{Pr:Compelexity}
    For a given function $f = \psi\Big(\sum_{k=1}\varphi_k(s_k)\Big)$, where $\varphi_k:\mathbb{R} \mapsto [-a,a]$ and $\psi: [-a,a] \mapsto [-b, b]$. Also, consider that the $m_k$-th derivation of  $\varphi_k$ and $\ell$-th derivation of $\psi$ exists and are bounded by factors $E_{k}$ and $D$, respectively. 
    The complexity of the encoders for computing the desired function $f$ over a network with $K$ nodes and the CP server is given by 
    \begin{align}
    \label{eq:FlopsEncoder}
    \#{\rm BOPs} =  \mathcal{O}\bigg(\sum_{k=1}^K\frac{\ln(\tfrac{E_kq}{2a\sqrt{2\pi}})}{\mathcal{W}\bigg({(2a{\rm e})}^{-1}\ln(\tfrac{E_kq}{2a\sqrt{2\pi}})\bigg)}\bigg), 
    \end{align}
     and for the decoder by
    \begin{align}
    \label{eq:FlopsDecoder}
    \#{\rm BOPs}  = \mathcal{O}\bigg(\frac{\ln(\tfrac{Dq}{2a\sqrt{2\pi}})}{\mathcal{W}\bigg({(2b{\rm e})}^{-1}\ln(\tfrac{Dq}{2a\sqrt{2\pi}})\bigg)}\bigg), 
    \end{align}
    for $q \gg 1$ and $\mathcal{W}(\cdot)$ denotes the Lambert $\mathcal{W}$ function.
\end{prop}
\begin{proof}
See Appendix~\ref{Ap:proofComp}. 
\end{proof}

From Proposition~\ref{Pr:Compelexity}, we note that the computational complexity of the encoder and the decoder of the SumComp coding are approximately $\mathcal{O}(K\ln{(q)})$ and $\mathcal{O}(\ln{(q)})$, respectively.  

Therefore, given the complexity of the SumComp coding and the original coding in  ChannelComp, we can conclude that the SumComp coding has an order of magnitude less computation costs than the coding included in ChannelComp, specifically when $q\gg1$.

In the next section, we evaluate the performance of the SumComp coding for computing different functions and compare it with the other methods, such as AirComp and the coding in ChannelComp. 

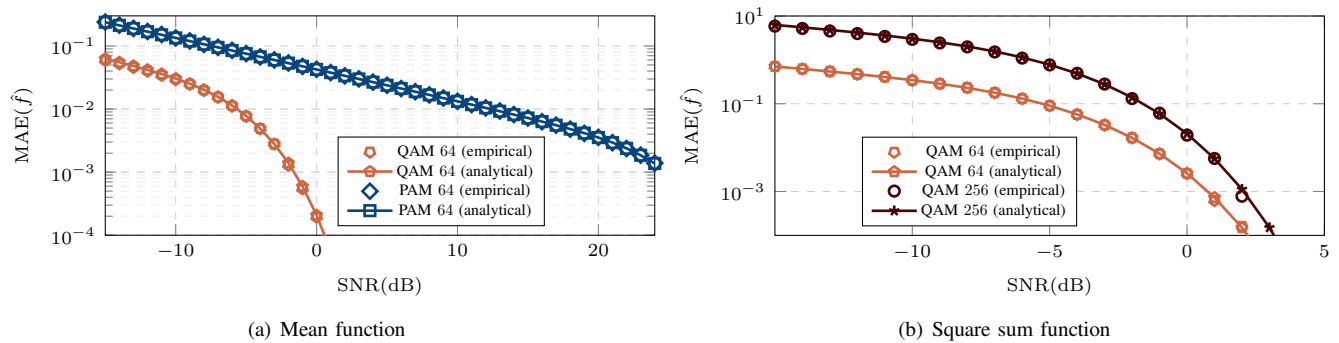
\begin{figure*}[!t]
\centering
\subfigure[Mean function]{
    \label{fig:QAMMean}
    \begin{tikzpicture} 
    \begin{axis}[
        xlabel={$\rm SNR (dB)$},
        ylabel={${\rm MAE}(\hat{f})$},
        label style={font=\scriptsize},
        tick label style={font=\scriptsize} , 
        width=0.49\textwidth,
        height=4.5cm,
        xmin=-15, xmax=24,
        ymin=1e-4, ymax=0.3,
         ymode = log,
       legend style={nodes={scale=0.6, transform shape}, at={(0.8,0.45)}}, 
        ymajorgrids=true,
        xmajorgrids=true,
        grid style=dashed,
        grid=both,
        grid style={line width=.1pt, draw=gray!15},
        major grid style={line width=.2pt,draw=gray!40},
    ]
     \addplot[only marks,
        color=burntsienna,
        mark=pentagon,
        mark options = {rotate = 36},
        line width=1pt,
        mark size=2pt,
        ]
    table[x=SNR,y=MAES1]
    {Data/MAEError.dat};
    \addplot[
        color=burntsienna,
        mark=pentagon,
        line width=1pt,
        mark size=2pt,
        ]
    table[x=SNR,y=MAET1]
    {Data/MAEError.dat};
         \addplot[only marks,
        color=darkcerulean,
        mark=square,
        mark options = {rotate = 45},
        line width=1pt,
        mark size=2pt,
        ]
    table[x=SNR,y=MAES2]
    {Data/MAEError.dat};
    \addplot[
        color=darkcerulean,
        mark=square,
        line width=1pt,
        mark size=2pt,
        ]
    table[x=SNR,y=MAET2]
    {Data/MAEError.dat};
    \legend{QAM $64$ (empirical), QAM $64$  (analytical),PAM $64$ (empirical), PAM $64$  (analytical) };
    \end{axis}
\end{tikzpicture}}\subfigure[Square sum function]{
    \label{fig:QAMNorm}
    \begin{tikzpicture} 
    \begin{axis}[
        xlabel={$\rm SNR (dB)$},
        ylabel={${\rm MAE}(\hat{f})$},
        label style={font=\scriptsize},
        tick label style={font=\scriptsize} , 
        width=0.49\textwidth,
        height=4.5cm,
        xmin=-15, xmax=5,
        ymin=1e-4, ymax=1e1,
        xtick={-10, -5, 0, 5, 10,15},
         ymode = log,
       legend style={nodes={scale=0.6, transform shape}, at={(0.55,0.45)}}, 
        ymajorgrids=true,
        xmajorgrids=true,
        grid style=dashed,
        grid=both,
        grid style={line width=.1pt, draw=gray!15},
        major grid style={line width=.2pt,draw=gray!40},
    ]
     \addplot[only marks,
        color=burntsienna,
        mark=pentagon,
        mark options = {rotate = 36},
        line width=1pt,
        mark size=2pt,
        ]
    table[x=SNR,y=MAES4]
    {Data/MAEError.dat};
    \addplot[
        color=burntsienna,
        mark=pentagon,
        line width=1pt,
        mark size=2pt,
        ]
    table[x=SNR,y=MAET4]
    {Data/MAEError.dat};
     \addplot[only marks,
        color=bulgarianrose,
        mark=o,
        line width=1pt,
        mark size=2pt,
        ]
    table[x=SNR,y=MAES3]
    {Data/MAEError.dat};
    \addplot[
        color=bulgarianrose,
        mark=star,
        line width=1pt,
        mark size=2pt,
        ]
    table[x=SNR,y=MAET3]
    {Data/MAEError.dat};
    \legend{QAM $64$ (empirical), QAM $64$  (analytical),QAM $256$ (empirical), QAM $256$  (analytical)};
    \end{axis}
\end{tikzpicture}

}
  \caption{ Monte Carlo evaluation of MAE of computing the mean and square norm function for $5\times10^4$ trials versus analytical results (Proposition \ref{Pr:MAE}) for digital modulation. Figure \ref{fig:QAMMean} shows both empirical and analytical MAE for QAM modulations and PAM modulation both of order  $q = 64$ in computing arithmetic mean function. Also, Figure \ref{fig:QAMNorm} shows both empirical and analytical MAE for QAM modulations of order $q = \{64, 256\}$ for computing the norm function. }
  \label{fig:SimNomo}
\end{figure*}

\section{Numerical Results}
\label{sec:numerical_results}

In this section, we assess the performance of the SumComp coding under various digital modulation schemes and corroborate the theoretical analysis through numerical results. In particular, we first evaluate the performance of SumComp coding with standard digital modulations, such as QAM and PAM, with different orders for computing summation function in terms of MSE metric. Moreover, we compare the output of empirical results from the simulation with the theoretical results from Section~\ref{sec:Performance}. We repeat this comparison for computing various functions in the class of nomographic functions in terms of the MAE metric. Afterward, we assess the performance of SumComp coding compared to the other existing methods. More precisely, the performance of the SumComp coding is compared to three different methods:
\begin{itemize}
\item ChannelComp original encoding: the optimization-based coding scheme proposed in \cite{Saeed2023ChannelComp} uses an optimization problem to obtain the digital modulation vectors.
\item AirComp method: the traditional AirComp method~\cite{goldenbaum2013harnessing,goldenbaum2014nomographic}, where nodes use analog modulation for communication. 
\item  Standard digital transmission: this scenario implements the computation method that relies on the naive OFDMA. This method ensures that each node is allocated unique frequency channels and prevents potential communication interference with more bandwidth consumption.
\end{itemize}
 Note that our comparison is limited to state-of-the-art coherent AirComp methods. Non-coherent methods, while viable, demand significantly higher communication resources compared to coherent counterparts, due to the absence of channel compensation.

Finally, we analyze the computational complexity of SumComp coding in comparison with the optimization-based coding in ChannelComp in terms of the number of BOPs. Our exploration and comparison over different metrics aim at providing a comprehensive understanding of the potential utility and performance of the SumComp coding.  Note that we use times sign, star, triangle, square, and pentagon markers to show the analytical results, plus signs, reverse triangles, diamonds, reverse pentagons, and circle markers for the empirical results.  Also, warm colors are reserved for QAM modulation, and cool colors are used for PAM modulation.{All the numerical experiment codes are provided on \href{https://github.com/SaeedRazavikia/SumComp-Coding-for-Digital-OAC.git}{Github}\footnote{\href{https://github.com/SaeedRazavikia/SumComp-Coding-for-Digital-OAC.git}{https://github.com/SaeedRazavikia/SumComp-Coding-for-Digital-OAC.git}}.}

\subsection{SumComp Coding with QAM \& PAM Modulations}

\begin{figure*}
\centering
\subfigure[Arithmetic mean]{
 \label{fig:QunNMSE(a)}
    \begin{tikzpicture} 
    \begin{axis}[
        xlabel={$\rm SNR~(dB)$},
        ylabel={NMSE},
        label style={font=\scriptsize},
        legend cell align={left},
        tick label style={font=\scriptsize} , 
        width=0.48\textwidth,
        height=4.5cm,
        xmin=-15, xmax=19,
        ymin=2e-5, ymax=1,
        ymode = log,
       legend style={nodes={scale=0.55, transform shape}, at={(0.98,0.98)}}, 
        ymajorgrids=true,
        xmajorgrids=true,
        grid style=dashed,
        grid=both,
        grid style={line width=.1pt, draw=gray!15},
        major grid style={line width=.2pt,draw=gray!40},
    ]
    \addplot[
        color=eggplant,
        mark=o,
        line width=1pt,
        mark size=2pt,
        ]
    table[x=SNR,y=ChannelSumA]
    {Data/NMSEerror.dat};
    \addplot[
        color=darklavender,
        mark=diamond,
        line width=1pt,
        mark size=2.5pt,
        ]
    table[x=SNR,y=ChannelCompA]
    {Data/NMSEerror.dat};
     \addplot[
        color=bluegray,
        mark=square,
        line width=1pt,
        mark size=2pt,
        ]
    table[x=SNR,y=OATA]
    {Data/NMSEerror.dat};
    \addplot[
        color=cadmiumgreen,
        mark=star,
        line width=1pt,
        mark size=2pt,
        ]
    table[x=SNR,y=OFDMA]
    {Data/NMSEerror.dat};
    \legend{SumComp coding $\sum$, ChannelComp~\cite{Saeed2023ChannelComp}  $\sum $,AirComp$\sum$, OFDMA$\sum$};
    \end{axis}
\end{tikzpicture}
}   \subfigure[Geometric mean]{
 \label{fig:QunNMSE(b)}
  \begin{tikzpicture} 
    \begin{axis}[
        xlabel={$\rm SNR~(dB)$},
        ylabel={NMSE},
        label style={font=\scriptsize},
        legend cell align={left},
        tick label style={font=\scriptsize} , 
        width=0.48\textwidth,
        height=4.5cm,
        xmin=-15, xmax=19,
        ymin=5e-5, ymax=10,
        ymode = log,
       legend style={nodes={scale=0.55, transform shape}, at={(0.98,0.98)}}, 
        ymajorgrids=true,
        xmajorgrids=true,
        grid style=dashed,
        grid=both,
        grid style={line width=.1pt, draw=gray!15},
        major grid style={line width=.2pt,draw=gray!40},
    ]
    \addplot[
        color=eggplant,
        mark=o,
        line width=1pt,
        mark size=2pt,
        ]
    table[x=SNR,y=ChannelSumG]
    {Data/NMSEerror.dat};
    \addplot[
        color=darklavender,
        mark=diamond,
        line width=1pt,
        mark size=2.5pt,
        ]
    table[x=SNR,y=ChannelCompG]
    {Data/NMSEerror.dat};
     \addplot[
        color=bluegray,
         mark=square,
        line width=1pt,
        mark size=2pt,
        ]
    table[x=SNR,y=OATG]
    {Data/NMSEerror.dat};
    \addplot[
        color=cadmiumgreen,
        mark=star,
        line width=1pt,
        mark size=2pt,
        ]
    table[x=SNR,y=OFDMG]
    {Data/NMSEerror.dat};
    \legend{SumComp coding$\prod$, ChannelComp~\cite{Saeed2023ChannelComp}  $\prod$,AirComp $\prod$,OFDMA $\prod$};
    \end{axis}
\end{tikzpicture}

}
  \caption{Performance comparison between SumComp coding, coding in ChannelComp, AirComp, and OFDMA in terms of NMSE  averaged over $N_s =5\times10^4$  Monte Carlo trials, when values of the function to be computed are originally quantized. The input values are set to $s_k=\{1, 2, \ldots, 64\}$  for the arithmetic mean, and $s_k=\{1, \ldots, 8\}$ for the geometric mean functions. Specifically, the desired functions are $f_1 = \sum_{k=1}^{10}s_k/K$,$f_2 = (\prod_{k=1}^4s_k)^{\tfrac{1}{K}}$.} 
  \label{fig:QunNMSE}  
\end{figure*}
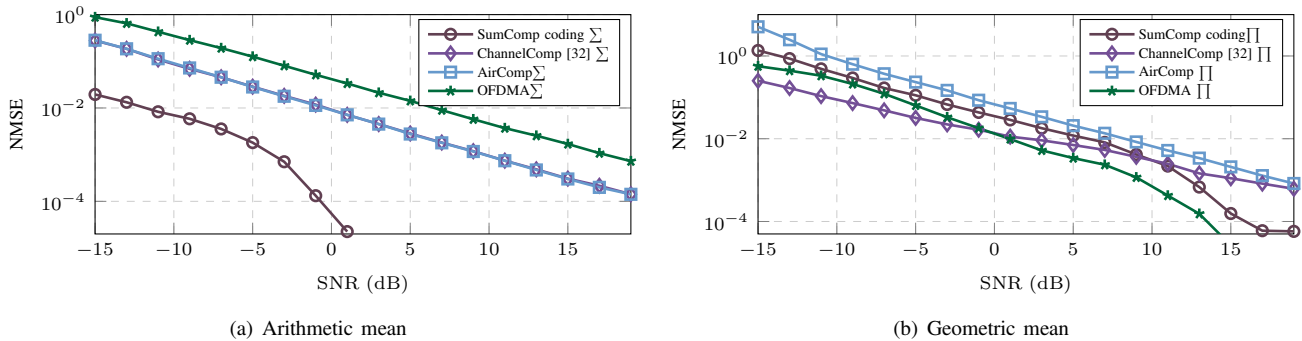

First, we analyze the performance of $\mathscr{E}_{k}$ encoders and $\mathscr{D}$ decoder for computing the summation function $f = \sum_{k=1}^{K}s_k$ over the MAC with $K=100$ nodes. The input data are generated in the same order as the modulations, i.e., $s_k \in \mathbb{Z}_{q}$ for $k\in [K]$, where $q$ is the order of modulation.  We consider two cases: nodes use QAM with $q=\{16,64,256\}$; and  PAM with $q=\{16,32,64\}$ over Gaussian channel for different levels of SNR, which is defined as ${\rm SNR}:=10\log(\sum_{i \in \mathcal{X}}|x_i|^2/q\sigma^2)$. We consider $5\times 10^4$ Monte Carlo trials.

In Figure~\ref{fig:SumSim}, we show the performance of the SumComp coding in terms of the MSE error when employed in different orders of QAM and PAM modulations. We also depict the analytical value of MSE in Proposition~\ref{Pr:MSE} for comparison with the outage of empirical error values. Figure~\ref{fig:SUMQAM} shows the empirical MSE error of the QAM of order $q$ where $q\in \{16,64,256\}$ with the triangle, pentagon, and circle markers, respectively. Similarly, we show the empirical MSE for the PAM of order $q$, where $q\in \{16,32,64\}$ with the plus sign, triangle, and square markers, respectively in Figure~\ref{fig:SUMPAM}. We note that the analytical MSE predicts well the average empirical error, as depicted in Figure~\ref{fig:SumSim}. Moreover, by increasing the order of the modulations, the MSE  values are higher over all ranges of SNR, which is expected because higher possible error values can occur due to the large size of the input field  $\mathbb{Z}_q$. This observation is also consistent with the analytical value of the  MSE in Proposition~\ref{Pr:MSE}.

\subsection{SumComp Coding for Nomographic Function}

 We consider nomographic functions and check the performance of the SumComp coding for two cases: the arithmetic mean and square sum functions (square Euclidean norm), as given in Examples~\ref{ex:Arithmetic} and \ref{ex:Euclidean}, respectively.  The computation is performed over a Gaussian MAC for $K=100$ nodes. In the case of arithmetic mean, the input data of node $k$, i.e., $s_k$ is uniformly random generated from $\mathbb{Z}_q$, where $q$ is set to be $64$ for both PAM and QAM modulations.  Then, in the next case, we repeat the experiment with QAM modulation with low and high order modulation, i.e., $q = 64$ and $256$. Moreover, the input data of node $k$, $s_k$, is generated uniformly random from $\{1,\ldots, \sqrt{q}\}$ to make $c_k = \varphi(s_k) \in \mathbb{Z}_{q}$.

Figure~\ref{fig:SimNomo} shows the performance of the SumComp coding for computing the arithmetic mean and the square sum functions in terms of MAE for different SNR values. In Figure~\ref{fig:QAMMean},  the MAE of computing arithmetic mean functions are depicted for different SNRs for nodes using either QAM or PAM modulations. We observe that the proposed upper bound in Proposition~\ref{Pr:MAE} approximates well the behavior of MAE curves, and the empirical results are consistent with Proposition~\ref{Pr:MAE}. Notably, QAM outperforms the PAM  thanks to its power-efficient constellation diagram. Indeed, the mean energy per transmitted symbol of PAM and QAM of order $q$ are $(q^2-1){A_s}^2/12$ and $(q-1){A_s}^2/6$, respectively, where $A_s$ denotes the distance between the constellation points.  Thus, the constellation diagram of QAM can approximately provide $10\log{(q+1)/2}$ SNR gain for the estimation of the received signal compared to PAM.  
 
Similarly, Figure~\ref{fig:QAMNorm} shows the MAE error for computing the square sum function where nodes use QAM modulation with order $q = 64$ and $=256$. Similar to Figure~\ref{fig:SumSim},  increasing the order of modulation results in a higher MAE error.  Similar to the arithmetic mean, the empirical results of MAE curves for the squared sum function are consistent with the upper bound proposed in Proposition~\ref{Pr:MAE}.

\subsection{SumComp Coding vs ChannelComp, AirComp, and OFDMA}

Now, we compare the SumComp coding with three methods: ChannelComp, AirComp, and OFDMA methods. The functions to compute are the arithmetic mean and geometric mean in terms of the normalized MSE (NMSE) metric, which is defined as 
${\rm NMSE}:= \sum_{j=1}^{N_s}|f_j - \hat{f}_j|^2/{N_s|f_j|},$
where $N_s$ denotes the number of Monte Carlo trials,  $f_j$ denotes the value of the desired function we wish to compute, and $\hat{f}_j$ is the estimated value of $f_j$ for $j\in [N_s]$. 

Figure~\ref{fig:QunNMSE(a)} shows the NMSE error for computing the arithmetic mean function. We note that SumComp coding outperforms all the other methods for computing the summation function even at the low SNR regime (less than $0$ dB). Specifically, for SNR around $-5$ dB, SumComp coding shows at least a $15$ dB improvement compared to the other methods. This significant improvement in the performance of SumComp coding is mainly attributed to the power-efficient constellation diagram of QAM compared to PAM. We note that the apparent similarity in performance between ChannelComp and AirComp in this figure is noteworthy. This resemblance arises from the specific characteristics of the summation function used in these scenarios. In the case of ChannelComp, the scheme employs the PAM modulation, while AirComp utilizes the analog AM modulation. However, when the input data for AirComp is discretized, the analog AM modulation effectively becomes PAM, thus aligning it with the ChannelComp method.

We repeat this experiment for the geometric mean in Figure~\ref{fig:QunNMSE(b)}. We observe that SumComp coding exhibits superior performance compared to ChannelComp and AirComp in the high SNR regime. In particular, for SNR greater than $12$ dB, SumComp coding NMSE has superiority over ChannelComp and AirComp, and it keeps increasing the performance for higher SNR until we observe more than $10$ dB improvement at SNRs around $19$ dB. Moreover, ChannelComp and OFDMA work better than the other methods for the low SNR regime.

Therefore, the results show that SumComp coding performs better in NMSE than other methods from the literature in computing the arithmetic and geometric mean, owing to the beneficial intersection of all nodes' constellation points and power-efficient QAM modulation. Additionally, the results imply that  SumComp coding exceeds the conventional AirComp approach across a broad range of SNR values, exhibiting notable superiority, particularly in the high SNR regime compared to both ChannelComp and AirComp.

\begin{figure*}[!t]
\centering
\subfigure[Encoder $q = 4$]{
    \label{fig:FlopsK}
    \begin{tikzpicture} 
    \begin{axis}[
        xlabel={$K$},
        ylabel={\# BOPs},
        label style={font=\scriptsize},
        tick label style={font=\scriptsize} , 
        width=0.25\textwidth,
        height=4cm,
        xmin=1, xmax=47,
        ymin=10, ymax=10^19,
         ymode = log,
       legend style={nodes={scale=0.55, transform shape}, at={(0.8,0.95)}}, 
        ymajorgrids=true,
        xmajorgrids=true,
        grid style=dashed,
        grid=both,
        grid style={line width=.1pt, draw=gray!15},
        major grid style={line width=.2pt,draw=gray!40},
    ]
     \addplot[
        color=eggplant,
        mark=o,
        line width=1pt,
        mark size=1.5pt,
        ]
    table[x=K,y=EncodeSumComp]
    {Data/ComplexityvsK.dat};
    \addplot[
        color=darklavender,
        mark=diamond,
        line width=1pt,
        mark size=2pt,
        ]
    table[x=K,y=EncodeChannelComp]
    {Data/ComplexityvsK.dat};
    \end{axis}
\end{tikzpicture}}\subfigure[Encoder $K=10$]{
    \label{fig:Flopsq}
    \begin{tikzpicture} 
    \begin{axis}[
        xlabel={$q$},
        label style={font=\scriptsize},
        tick label style={font=\scriptsize} , 
        width=0.25\textwidth,
        height=4cm,
        xmin=2, xmax=128,
        ymin=1, ymax=5*10^27,
         ymode = log,
       legend style={nodes={scale=0.55, transform shape}, at={(0.8,0.95)}}, 
        ymajorgrids=true,
        xmajorgrids=true,
        grid style=dashed,
        grid=both,
        grid style={line width=.1pt, draw=gray!15},
        major grid style={line width=.2pt,draw=gray!40},
    ]
     \addplot[
        color=eggplant,
        mark=o,
        line width=1pt,
        mark size=1.5pt,
        ]
    table[x=q,y=EncodeSumComp]
    {Data/Complexityvsq.dat};
    \addplot[
        color=darklavender,
        mark = diamond,
        line width=1pt,
        mark size=2pt,
        ]
    table[x=q,y=EncodeChannelComp]
    {Data/Complexityvsq.dat};
    \end{axis}
\end{tikzpicture}

}\subfigure[Decoder $q = 4$]{
    \label{fig:FlopsKDe}
    \begin{tikzpicture} 
    \begin{axis}[
        xlabel={$K$},
        label style={font=\scriptsize},
        tick label style={font=\scriptsize} , 
        width=0.25\textwidth,
        height=4cm,
        xmin=1, xmax=47,
        ymin=1, ymax=10^28,
         ymode = log,
       legend style={nodes={scale=0.55, transform shape}, at={(0.8,0.95)}}, 
        ymajorgrids=true,
        xmajorgrids=true,
        grid style=dashed,
        grid=both,
        grid style={line width=.1pt, draw=gray!15},
        major grid style={line width=.2pt,draw=gray!40},
    ]
     \addplot[
        color=eggplant,
        mark=o,
        line width=1pt,
        mark size=1.5pt,
        ]
    table[x=K,y=DecodeSumComp]
    {Data/ComplexityvsK.dat};
    \addplot[
        color=darklavender,
        mark=diamond,
        line width=1pt,
        mark size=2pt,
        ]
    table[x=K,y=DecodeChannelComp]
    {Data/ComplexityvsK.dat};
    \end{axis}
\end{tikzpicture}}\subfigure[Decoder $K=10$]{
    \label{fig:FlopsqDe}
    \begin{tikzpicture} 
    \begin{axis}[
        xlabel={$q$},
        label style={font=\scriptsize},
        tick label style={font=\scriptsize} , 
        width= 0.25\textwidth,
        height=4cm,
        xmin=2, xmax=128,
        ymin=10, ymax=10^21,
         ymode = log,
       legend style={nodes={scale=0.6, transform shape}, at={(0.98,0.45)}}, 
        ymajorgrids=true,
        xmajorgrids=true,
        grid style=dashed,
        grid=both,
        grid style={line width=.1pt, draw=gray!15},
        major grid style={line width=.2pt,draw=gray!40},
    ]
     \addplot[
        color=eggplant,
        mark=o,
        line width=1pt,
        mark size=1.5pt,
        ]
    table[x=q,y=DecodeSumComp]
    {Data/Complexityvsq.dat};
    \addplot[
        color=darklavender,
        mark = diamond,
        line width=1pt,
        mark size=2pt,
        ]
    table[x=q,y=DecodeChannelComp]
    {Data/Complexityvsq.dat};
    \legend{SumComp coding, ChannelComp~\cite{Saeed2023ChannelComp}};
    \end{axis}
\end{tikzpicture}

}

  \caption{ Computational complexity for both encoder and decoder of SumComp coding and coding in ChannelComp in terms of BOPs. Figures \ref{fig:FlopsK} and \ref{fig:FlopsKDe} show the encoder's and decoder's complexity when $q=4$ for different numbers of node $K$, respectively.  Similarly, Figures \ref{fig:Flopsq} and \ref{fig:FlopsqDe} show the encoder's and decoder's complexity when $K=10$ for different quantization levels $q$, respectively. }
  \label{fig:Complixity}
\end{figure*}
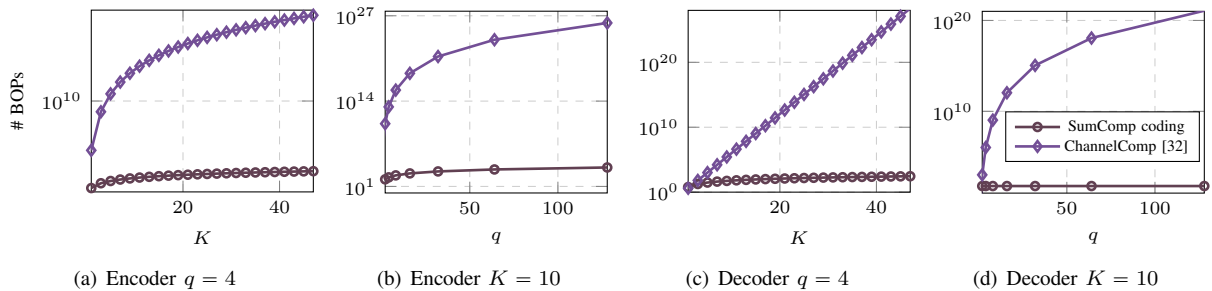

\subsection{Complexity performance}

Finally, we compare the computational complexity of the SumComp coding with the optimization-based method by ChannelComp~\cite{razavikia2023computing}.  

We show the order of complexity of ChannelComp and SumComp coding in Figure~\ref{fig:Complixity} for the summation function, where the number of constraints is reduced to $K^2q^2$, and because of symmetry, the dimension of variables becomes $n = q$. Therefore, the overall computational complexity for both the encoder and decoder of the summation function becomes $\mathcal{O}(K^8q^{8.5})$.  For SumComp coding, the required number of operations for the encoder is simply $\mathcal{O}(K)$, and for computing, the decoder is $\mathcal{O}(1)$ because $\psi(y) = y$. In Figure~\ref{fig:Complixity}, the complexity of  ChannelComp and SumComp coding are depicted for different numbers of nodes and quantization levels, respectively.  We observe that the SumComp coding can significantly reduce the huge complexity of the optimization solution for ChannelComp, while it shows similar performance in the computing class of nomographic functions.

\section{Conclusions}
\label{sec:conclusions}

In conclusion, this study identified the potential to enhance the ChannelComp coding procedure by merging communication and computation, but the inherent complexity poses challenges concerning its broad optimization applicability.  To mitigate these challenges, we introduced the SumComp coding, an innovative and straightforward coding scheme based on the ring of integers that allows for the implementation of digital modulations. The SumComp coding exhibited remarkable compatibility across several digital modulations, such as  QAM and PAM, to name a few. In addition, we analyzed the MSE for SumComp coding in the computation of the arithmetic mean function and established an upper bound on the MAE for various nomographic functions.

An important conclusion from our work is the demonstrable superiority of SumComp coding over traditional AirComp methods. The empirical results substantiated this claim and implied significant advancement in digital modulation for AirComp. Specifically, for computing arithmetic and geometric mean in numerical results,  SumComp coding shows around $10$ dB improvements in terms of normalized MSE for low-noise scenarios.  The proposed methodology of SumComp coding not only overcomes the computational complexity of the original coding scheme of ChannelComp but also paves the way for further exploration of digital wireless computation.

The potential of SumComp coding inspires many avenues for further study. Here, we summarize the most promising trajectories. 
\begin{itemize}
    \item \textbf{Channel code enhancement:} develop a SumComp channel code resistant to noise, such that optimizing the transmission number and computation error becomes possible.
    \item \textbf{Expanding function computation:} while SumComp coding successfully computes summation functions, considering other ring algebraic structures, such as the polynomial ring and the quotient ring, could allow for a wider range of functions computed over the air. 
    \item \textbf{MIMO extension for matrix computation:} progressing from a single narrowband antenna system to a MIMO system enables vector-based computations, paving the way for more applications, such as matrix computation and federated learning.
    \item \textbf{Machine learning applications:} we envision that SumComp coding can be applied and significantly enhance applications in edge federated learning.
    \item \textbf{Real-world implementation:} evaluating the SumComp coding with real-world experiments to demonstrate SumComp's feasibility in practical wireless communication systems. 
    
\end{itemize}


\appendix
\subsection{Proof of Proposition \ref{Pr:Group}}\label{Ap:GroupProof}
For $\mathcal{S}_{q_1,q_2}^{\rho}$ to be a group under the operation of summation, it needs to satisfy the \textit{closure}, \textit{associativity}, \textit{identity element}, and \textit{inverse element} property of a group.  Each property can be proved as follows. First, note that $\mathcal{S}_{q_1,q_2}^{\rho}$ is non-empty because it contains at least the identity element $(0,0)$.

\textit{Closure}: Let us suppose there are two elements $(c_1, x_1)$ and $(c_2, x_2)$ in $\mathcal{S}_{q_1,q_2}^{\rho}$, where $c_1,c_2\in \mathbb{Z}$, and  $x_1,x_2\in \mathbb{Z}[\rho]$. Then, we need to prove that their sum, which is $(c_1+c_2, x_1+x_2)$, also belongs to $\mathcal{S}_{q_1,q_2}^{\rho}$. To this end, we have
\begin{align}
  c_1 & = \mathfrak{Re}(\mathcal{G}_{\rho}^{-1}(x_1))q_1 + \mathfrak{Im}(\mathcal{G}_{\rho}^{-1}(x_1))q_2,\\
  c_2 & = \mathfrak{Re}(\mathcal{G}_{\rho}^{-1}(x_2))q_1 + \mathfrak{Im}(\mathcal{G}_{\rho}^{-1}(x_2))q_2.
\end{align}
By adding up both sides, we have
\begin{align}
\nonumber
c_1 + c_2 = &\mathfrak{Re}(\mathcal{G}_{\rho}^{-1}(x_1 +x_2))q_1\\ 
&+ \mathfrak{Im}(\mathcal{G}_{\rho}^{-1}(x_1+x_2))q_2 \in \mathcal{S}_{q_1,q_2}^{\rho}.
\end{align}
Therefore, $(c_1 + c_2,x_1+x_2) \in \mathcal{S}_{q_1,q_2}^{\rho}$ and the closure property is satisfied.  \textit{Associativity}: This property holds because the operation of addition is associative in $\mathbb{R}^2$ and $\mathbb{R}$.

\textit{Identity element}: The identity element is $(0, 0)$ since for any element $(c, x) \in \mathcal{S}_{q_1,q_2}^{\rho}$, we have $(0+c, 0+x) = (c, x)$.

\textit{Inverse element}: For every element $(c, x) \in \mathcal{S}_{q_1,q_2}^{\rho}$, the inverse is given by $(-c, -x)$ because $c + (-c) = 0$ and $x + (-x) = 0$.

Since all group axioms are satisfied, we conclude that $\mathcal{S}_{q_1,q_2}^{\rho}$ is a group under the operation of summation.
\subsection{Proof of Proposition \ref{Pr:MSE}}\label{Ap:MSEProof}
To compute the MSE for the arithmetic sum function, i.e., $f= \sum_{k=1}^Ks_k$ for $s_k\in \mathbb{Z}_q$, we first need to compute the probability of the error for each individual value of constellation points. To this end, we have 
\begin{align}
\nonumber
 r & = \sum\nolimits_{k=1}^Kx_k + z =   \sum\nolimits_{k=1}^K\{\mathcal{G}_{\rho}\big(\mathcal{E}_{q_1,q_2}(s_k)\big)\}  + z, \\ \nonumber
 & = \sum_{k=1}^K \{s_k\mu_{1} + m_kq_2 +  (s_k\mu_{2} - m_kq_1){i}\rho\} + z,  \\ \nonumber
 & = \sum_{k=1}^K \{s_k\mu_{1} + m_kq_2 +  (s_k\mu_{2} - m_kq_1){i}\rho\} \\
 & \quad + z_1 + z_2\rho i, \label{eq:wtoen}
\end{align}
where $\mu_1q_1+ \mu_2q_2 = 1$, and $ z_1 + z_2\rho  \in \mathbb{C}$ is the representation of noise $z$ in the ring of $\mathbb{Z}[\rho]$. Then, using \eqref{eq:wtoen}, the decoder $\mathscr{D}$ yields 
\begin{align}
    \nonumber
  \hat{f} & =  \mathcal{D}_{q_1,q_2}\mathcal{G}^{-1}_{\rho}\mathcal{Q}_{\rho}(r) \\ \nonumber &=  \mathcal{D}_{q_1,q_2}\mathcal{G}^{-1}_{\rho}\Bigg(\sum\nolimits_{k=1}^K \{s_k\mu_{1} + m_kq_2\} + \mathcal{Q}_{\rho}(z_1)    \\\nonumber &  \qquad +   \sum\nolimits_{k=1}^K \{ (s_k\mu_{2} - m_kq_1){i} \rho  + \mathcal{Q}_{\rho}(z_2)\rho i \}\Bigg),\\ \nonumber
  & = \mathcal{D}_{q_1,q_2}\Bigg(\sum\nolimits_{k=1}^K \{s_k\mu_{1} + m_kq_2\} + \mathcal{Q}_{\rho}(z_1) \\ \nonumber
  &  \qquad +   \sum\nolimits_{k=1}^K \{ (s_k\mu_{2} - m_kq_1){i} + \mathcal{Q}_{\rho}(z_2)i \}\Bigg),\\ \nonumber
   & = q_1\times \sum\nolimits_{k=1}^K \{s_k\mu_{1} + m_kq_2\} + q_1\times\mathcal{Q}_{\rho}(z_1) + \\ \nonumber
  & \qquad  q_2\times \sum\nolimits_{k=1}^K \{ (s_k\mu_{2} - m_kq_1)  + q_2\times \mathcal{Q}_{\rho}(z_2) \}, \\ \nonumber
  & =  \sum_{k=1}^K \{s_k\mu_{1}q_1 + m_kq_2q_1\} +  \sum_{k=1}^K \{s_kq_2\mu_{2} - m_kq_1q_2\}  \\ \nonumber
  &+ q_1\mathcal{Q}_{\rho}(z_1) + q_2 \mathcal{Q}_{\rho}(z_2), \\ 
  & = f + q_1\mathcal{Q}_{\rho}(z_1) + q_2 \mathcal{Q}_{\rho}(z_2).
 \label{eq:noisedecom}
\end{align}
Hence, for the MSE, we can write
\begin{align}
\nonumber
\mathbb{E}\big\{|f - \hat{f}|_2^2\big\} & = \mathbb{E}\big\{|q_1\mathcal{Q}_{\rho}(z_1) + q_2 \mathcal{Q}_{\rho}(z_2)|^2\},\\ 
  & = q_1^2\mathbb{E}\big\{{|\underbrace{\mathcal{Q}_{\rho}(z_1)}_{:=e_{1}}|}^2\}  + q_2^2\mathbb{E}\big\{{|\underbrace{\mathcal{Q}_{\rho}(z_2)}_{:=e_{2}}|}^2\}, 
\end{align}
where the last equality comes from the fact that $z_{1}$ and $z_{2}$ are statistically independent random variables and the finite rectangular detector is symmetric along each coordinate, the induced coordinate
errors $e_1$ and $e_2$ are independent and zero-mean.  Since noise components $z_{1}$ and $z_{2}$ are independent random variables, we can independently treat the error of the real part $e_{1}$ and the imaginary part $e_{2}$ in \eqref{eq:noisedecom}. The operator $\mathcal{Q}_{\rho}$ in $1$D acts as the round function, and it maps the input values from $\mathbb{R}$ to its nearest neighbor integer value.  Consequently, variables $e_{1}$ and $e_{2}$ are integers corresponding to the modulation's error in-phase and quadrature components. Because of noise in every dimension, the receiver may add the discretized value of error $z_{1}$ (or $z_{2}$) from the $\mathbb{Z}$ to the estimated value $\hat{f}$. Then, the computation error in the real domain equals the absolute value of $e_{1}$ times the step size $q_1$. Similarly,  the computation error in the imaginary domain is $|e_{2}|\times q_2$.

In this regard, let $e_1$ and $e_2$ denote the detection errors along the
real and imaginary coordinates of the finite aggregate constellation
$\mathcal{X}^{(K)}$, respectively. Using the law of total expectation, we have
\begin{subequations}
\label{eq:Expe1e2}
\begin{align}
\mathbb{E}[e_1^2]
&=
\sum\nolimits_{j_1=0}^{M_1-1}
\mathbb{E}[e_1^2|s_1=j_1]\Pr[s_1=j_1],  \\
\mathbb{E}[e_2^2]
&=
\sum\nolimits_{j_2=0}^{M_2-1}
\mathbb{E}[e_2^2|s_2=j_2]\Pr[s_2=j_2],
\end{align}
\end{subequations}
where $s_1$ and $s_2$ denote the real and imaginary coordinate indices of the
noiseless aggregate constellation point, respectively. Under the assumption of
a uniform distribution over the induced aggregate constellation points, we have
$\Pr[s_1=j_1]=1/M_1$ and $\Pr[s_2=j_2]=1/M_2$\footnote{This is the marginal
probability distribution for the real and imaginary axes. Since the joint
distribution is uniform, its marginals are also uniform.}.

For the finite nearest-neighbor detector, the event $|e_1|\geq \ell_1$ occurs
when the noise crosses at least $\ell_1$ decision regions. For a given
coordinate index $j_1$, this can happen toward the right boundary if
$j_1\leq M_1-\ell_1-1$, or toward the left boundary if $j_1\geq \ell_1$.
Hence, after averaging over the uniform distribution of $s_1$, we obtain
\begin{align}
\Pr\{|e_1|\geq \ell_1\}
=
\frac{2(M_1-\ell_1)}{M_1}
Q\Big(\frac{(2\ell_1-1)d_1}{\sigma_1}\Big).
\end{align}
Similarly,
\begin{align}
\Pr\{|e_2|\geq \ell_2\}
=
\frac{2(M_2-\ell_2)}{M_2}
Q\Big(\frac{(2\ell_2-1)d_2}{\sigma_2}\Big).
\end{align}
Using the identity
\begin{align}
\mathbb{E}[e_i^2]
=
\sum_{\ell_i=1}^{M_i-1}
(2\ell_i-1)\Pr\{|e_i|\geq \ell_i\},
\quad i\in\{1,2\},\label{eqExpe1}
\end{align}
we obtain
\begin{subequations}
\label{eq:Expzet}
\begin{align*}
\mathbb{E}[e_1^2]
&=
2\sum\nolimits_{\ell_1=1}^{M_1-1}
(2\ell_1-1)
\Big(1-\frac{\ell_1}{M_1}\Big)
Q\Big(\frac{(2\ell_1-1)d_1}{\sigma_1}\Big),\\
\mathbb{E}[e_2^2]
&=
2\sum\nolimits_{\ell_2=1}^{M_2-1}
(2\ell_2-1)
\Big(1-\frac{\ell_2}{M_2}\Big)
Q\Big(\frac{(2\ell_2-1)d_2}{\sigma_2}\Big).
\end{align*}
\end{subequations}
Here, $d_1$ and $d_2$ are the in-phase and quadrature decision distances,
respectively, which are $d_1=d_2=1/2$ for Gaussian integers. Moreover,
$\sigma_1$ and $\sigma_2$ denote the standard deviations of the effective
noise along the two axes, which are $\sigma_1=\sigma$ and
$\sigma_2=\sigma/|\rho|$, respectively. Also, recall that $Q(x)$ is the
Gaussian $Q$ function.

Therefore, we can write
\begin{subequations}
\begin{align}
    \label{eq:E_q1}
    \mathbb{E}[e_{1}^2]
    &=
    \sum\nolimits_{\ell_1=1}^{M_1-1}
    \frac{\alpha_1(\ell_1)}{q_1^2}
    Q\Big(\frac{2\ell_1-1}{2\sigma}\Big),\\
    \label{eq:E_q2}
    \mathbb{E}[e_{2}^2]
    &=
    \sum\nolimits_{\ell_2=1}^{M_2-1}
    \frac{\alpha_2(\ell_2)}{q_2^2}
    Q\Big(\frac{(2\ell_2-1)|\rho|}{2\sigma}\Big),
\end{align}
\end{subequations}
where
\begin{subequations}
\begin{align}
\alpha_1(\ell_1)
&=
2q_1^2(2\ell_1-1)
\Big(1-\frac{\ell_1}{M_1}\Big),\\
\alpha_2(\ell_2)
&=
2q_2^2(2\ell_2-1)
\Big(1-\frac{\ell_2}{M_2}\Big).
\end{align}
\end{subequations}
Substituting \eqref{eq:E_q1} and \eqref{eq:E_q2} into the MSE expression gives
\[
{\rm MSE}(\hat f)=q_1^2\mathbb{E}[e_1^2]+q_2^2\mathbb{E}[e_2^2],
\]
which yields the expression in Proposition~\ref{Pr:MSE}. Therefore, the proof
is complete.
\subsection{Proof of Proposition \ref{Pr:MAE}}\label{Ap:MAEProof}

By recalling the definition of the ${\rm MSE}(\hat{f}) = \mathbb{E}[|f-\hat{f}|]$, we have
\begin{align}
\nonumber
    \mathbb{E}[|f-\hat{f}|] &=\mathbb{E}\Big[\big|\psi(\sum\nolimits_{k=1}^Ks_k)-\psi(\sum\nolimits_{k=1}^K\hat{s}_k)\big|\Big], \\ \nonumber &   \leq \mathbb{E}\bigg[\Big| w_{\psi}\big(|\sum\nolimits_{k=1}^Ks_k - \sum\nolimits_{k=1}^K\hat{s}_k|\big)\Big|\bigg],  \\ \nonumber
    & = \mathbb{E}\bigg[ w_{\psi}\Big(\big|\sum\nolimits_{k=1}^Ks_k - \sum\nolimits_{k=1}^K\hat{s}_k\big|\Big)\bigg], \\ & \leq  w_{\psi}\bigg(\mathbb{E}\Big[\Big|\sum\nolimits_{k=1}^Ks_k - \sum\nolimits_{k=1}^K\hat{s}_k\Big|\Big]\bigg), \label{eq:MSEbound11}  
\end{align}
where the final inequality represents the reversed Jensen's inequality \cite{mcshane1937jensen}, due to the concavity of $w_{\psi}$.
To obtain the expected value in the argument of function $w_{\psi}$, we use \eqref{eq:noisedecom} as follows 
\begin{align}
\nonumber
\mathbb{E}\Big[\Big|\sum_{k=1}^Ks_k - &\sum_{k=1}^K\hat{s}_k\Big|\Big]  = \mathbb{E}\big\{|q_1\mathcal{Q}_{\rho}(z_1) + q_2 \mathcal{Q}_{\rho}(z_2)|\}, \\ \label{eq:UpL1Error}
& \leq q_1  \mathbb{E}\big\{|\mathcal{Q}_{\rho}(z_1)|\} + q_2 \mathbb{E}\big\{|\mathcal{Q}_{\rho}(z_2)|\},
\end{align}
where the last inequality is due to the triangle inequality. By following a similar procedure from  \eqref{eq:Expe1e2} to \eqref{eqExpe1}, we can compute the expectation of the error terms as follows:
\begin{subequations}
\begin{align}
\mathbb{E}[e_{1}] & = 2\sum_{\ell_1=1}^{M_1-1} (\ell_1- \frac{\ell_1^2}{M_1})\zeta_1(\ell_1), \\
\mathbb{E}[e_{2}] & =2 \sum_{\ell_2=1}^{M_2-1} (\ell_2-\frac{\ell_2^2}{M_2}) \zeta_2(\ell_2). 
\end{align}
\end{subequations}
With further simplifications, we obtain the following terms:
\begin{subequations}
\label{eq:Probe2}
\begin{align}
    \mathbb{E}[e_{1}] & = \hspace{-2pt} 2 \hspace{-5pt}\sum_{\ell_1=1}^{M_1-1}\hspace{-5pt}\Big(1 - \frac{\ell_1}{M_1}\Big) Q\big(\frac{(2\ell_1-1)}{2\sigma}\big), \\
   \mathbb{E}[e_{2}] & = \hspace{-2pt} 2 \hspace{-5pt}\sum_{\ell_2=1}^{M_2-1}\hspace{-5pt}\Big(1 - \frac{\ell_2}{M_2}\Big) Q\big(\frac{(2\ell_2-1)|\rho|}{2\sigma}\big).
   \end{align}
\end{subequations}
Then, by substituting \eqref{eq:Probe2} into \eqref{eq:UpL1Error}, we obtain 
\begin{align}
\nonumber
\mathbb{E}\Big[\Big|\sum\nolimits_{k=1}^Ks_k - &\sum\nolimits_{k=1}^K\hat{s}_k\Big|\Big]  \leq q_1 \beta_1 +  q_2 \beta_2, 
\end{align}
where, $\beta_1$ and $\beta_2$ are given by  
\begin{subequations}
\begin{align}
    \beta_1 & = 2 \sum\nolimits_{\ell_1=1}^{M_1-1}\Big(1 - \frac{\ell_1}{M_1}\Big) Q\big(\frac{(2\ell_1-1)}{2\sigma}\big), \\
   \beta_2 & =  2 \sum\nolimits_{\ell_2=1}^{M_2-1}\Big(1 - \frac{\ell_2}{M_2}\Big) Q\big(\frac{(2\ell_2-1)|\rho|}{2\sigma}\big).
   \end{align}
\end{subequations}
Finally, recalling \eqref{eq:MSEbound11} and the fact that $w_{\psi}$ is a strictly increasing function, we obtain
\begin{align}
\nonumber
\mathbb{E}[|f-\hat{f}|]  \leq  w_{\psi}\Big(q_1 \beta_1 +  q_2 \beta_2  \Big).
\end{align}
Hence, we conclude the proof of Proposition \ref{Pr:MAE}.

\subsection{Proof of Proposition \ref{Pr:Compelexity}}\label{Ap:proofComp}

To analyze the complexity of the encoders $\mathscr{E}_k$, we note that $\mathcal{E}_k$ and $\mathcal{G}_{\rho}$ are the same for all the nodes, and they only involve four and two operations, respectively.   Hence, their complexities are $\mathcal{O}(1)$, and they do not change the complexity order of the computations. Similarly, for corresponding decoder parts, i.e., $\mathcal{D}$, $\mathcal{G}_{\rho}^{-1}$, and have complexity of $\mathcal{O}(1)$. Furthermore, the complexity of the rounding operator $\mathcal{Q}$ is also $\mathcal{O}(1)$ because it involves only basic arithmetic operations such as truncation and addition.

Therefore, we must compute the complexity only for the $\varphi_k$ and $\psi$ functions. To this end, we use the Taylor Series expansion to compute a general function with a complexity of $\mathcal{O}(n)$, where $n$  represents the number of terms in the series.  Because the required accuracy for given functions $\varphi_k$ is at most $1/q$, we consider Taylor's remainder a computation error~\cite{folland1990remainder}. Hence, we can write the following for computing $\varphi_k$ around point $s \in [-a, a]$:  
\begin{align}
   \frac{|\varphi_k^{(m_k)}(s)|}{(m_k+1)!}(2a)^{(m_k+1)} &  \leq \frac{(2a)}{q},
\end{align}
where  $\varphi_k^{(m_k)}$ denotes the $m_k$-th derivation of function $\varphi_k$. Then, using the fact that $|\varphi_k^{(m_k)}(s)|\leq E_k$, we have
\begin{align}
   \frac{E_k}{(m_k+1)!}(2a)^{(m_k+1)} &  \leq \frac{2a}{q}, \\
   \frac{E_kq}{2a} \leq \frac{(m_k+1)!}{(2a)^{(m_k+1)}}.
\end{align}
Then,  by using Stirling approximation for $q \gg 1 $ and doing manipulation \cite{amenyou2018properties}, we reach  
\begin{align}
   m_k \geq \Bigg\lceil \frac{\ln(\tfrac{E_kq}{2a\sqrt{2\pi}})}{\mathcal{W}\bigg({(2a{\rm e})}^{-1}\ln(\tfrac{E_kq}{2a\sqrt{2\pi}})\bigg)}\Bigg\rceil,
\end{align}
in which $\mathcal{W}(\cdot)$  is Lambert function defined as $\mathcal{W}(x) := z$ for $z{\rm e}^z = x$~\cite{corless1996lambert}. The number of operations must be calculated to compute the function $\varphi_k$. As a result, for computing all the encoders, we need to add them up, which leads to the number in \eqref{eq:FlopsEncoder}.  Following a similar procedure, we can show the complexity of computing the post-processing function $\psi$. Therefore, we conclude the proof.

\subsection{Acknowledgment}

The authors would like to thank Juan Gonzalez for identifying a minor issue concerning the treatment of the tail points in the calculation within an earlier proof of Proposition~\ref{Pr:MSE}.

\bibliographystyle{IEEEtran}
\bibliography{IEEEabrv,Ref}

\end{document}